\documentclass[prd,aps,twocolumn,a4paper,floatfix,nofootinbib]{revtex4-2}

\usepackage{graphicx}
\usepackage{dcolumn}
\usepackage{bm}
\usepackage{graphicx}
\usepackage{subfigure}
\usepackage{array}
\usepackage{booktabs}
\usepackage {xcolor}
\usepackage{amssymb}
\usepackage{pifont}
\usepackage{multirow}
\usepackage{diagbox}
\usepackage{comment}
\usepackage{xcolor}
\usepackage{enumerate}
\usepackage{booktabs}
\usepackage[normalem]{ulem}
\usepackage[utf8]{inputenc}
\usepackage{hyperref}
\usepackage{units}
\usepackage{enumerate}
\usepackage{amsmath}
\hypersetup{
    colorlinks = true,
    linkcolor = {blue},
    citecolor = {blue},
    urlcolor = {blue},
    linkbordercolor = {white},
    }
\usepackage{orcidlink}
\usepackage{float}

\begin{document}

\title{A machine-learning classifier for the postmerger remnant of binary neutron stars}

\author{Anna Puecher$^{1}$\orcidlink{0000-0003-1357-4348}}

\author{Tim Dietrich$^{1,2}$\orcidlink{0000-0003-2374-307X}}

\affiliation{${}^1$Institut f\"{u}r Physik und Astronomie, Universit\"{a}t Potsdam, Haus 28, Karl-Liebknecht-Str. 24/25, 14476, Potsdam, Germany}
\affiliation{${}^2$Max Planck Institute for Gravitational Physics (Albert Einstein Institute), Am M\"uhlenberg 1, Potsdam 14476, Germany}

\date{\today}

\begin{abstract}
    Knowing the kind of remnant produced after the merger of a binary neutron star system, e.g., if a black hole forms or not, would not only shed light on the equation of state describing the extremely dense matter inside neutron stars, but also help understand the physical processes involved in the postmerger phase. Moreover, in the event of a gravitational-wave detection, predicting the presence of a neutron star remnant is crucial in order to advise potential electromagnetic follow-up campaigns. In this work, we use Gradient Boosted Decision Trees and publicly available data from numerical-relativity simulations to construct a classifier that predicts the outcome of binary neutron star mergers, based on the binary's parameters inferred from gravitational-wave inspiral signals: total mass, mass-weighted tidal deformability, mass ratio, and effective inspiral spin. Employing parameters that can be estimated from the inspiral part of the signal only allows us to predict the remnant independently on the detection of a postmerger gravitational-wave signal. We build three different classifiers to distinguish between various potential scenarios, we estimate their accuracy and the confidence of their predictions.
    Finally, we apply the developed classifiers to real events data, finding that GW170817 most likely led to the formation of a hypermassive neutron star, while GW190425 to a prompt collapse to a black hole.
\end{abstract}

\maketitle

\section{Introduction}
\label{sec:intro}

Binary neutron star (BNS) mergers allow us to study matter at extremely high densities (see, e.g., Refs.~\cite{Baiotti:2016qnr,Dietrich:2020eud,Radice:2020ddv} for a review). One of the main tools to investigate these phenomena is the gravitational-wave (GW) signal emitted during the entire coalescence, since the GWs morphology strongly depends on the equation of state (EoS) of the matter forming the neutron stars (NSs) in the system {(see, e.g., Refs.~\cite{Hinderer:2009ca,Vines:2011ud,Baiotti:2016qnr,Agathos:2015uaa, Damour:2012yf,Carson:2019xxz} and references therein). Advanced LIGO~\cite{LIGOScientific:2014pky} and Advanced Virgo~\cite{VIRGO:2014yos} already observed two such events, GW170817~\cite{LIGOScientific:2017vwq} and GW190425~\cite{LIGOScientific:2020aai}, and more detections are foreseen in the upcoming years, both with current and especially with future-generation detectors, such as the Einstein Telescope~\cite{Punturo:2010zz,Maggiore:2019uih,LIGOScientific:2016wof,Hild:2010id} and the Cosmic Explorer~\cite{Reitze:2019iox,Evans:2021gyd,Srivastava:2022slt}.

From the analysis of the GW signal emitted during the inspiral of a BNS system, we can measure some of the NSs' properties, such as their mass, spin, and tidal deformability~\cite{Dietrich:2020eud}, which already allows us to place constraints on the EoS of supranuclear-dense matter~\cite{Dietrich:2020eud,Chatziioannou:2020pqz, LIGOScientific:2018hze, LIGOScientific:2018cki,Iacovelli:2023nbv}. Moreover, additional knowledge about the EoS of matter at higher densities and temperatures can be inferred by the potential detection of a postmerger signal~\cite{Baiotti:2016qnr,Bauswein:2013jpa, Koppel:2019pys,Bauswein:2015vxa,Bauswein:2019ybt, Sarin:2020gxb,Breschi:2021xrx,Puecher:2022oiz,Bauswein:2020aag,Kashyap:2021wzs}, which is expected to become feasible with next-generation detectors~\cite{Branchesi:2023mws,Maggiore:2019uih,Reitze:2019iox,Evans:2021gyd,Srivastava:2022slt,LIGOScientific:2017fdd,LIGOScientific:2018urg}.

The final fate of the system depends on the EoS, the binary's total mass, and its other parameters. Each EoS predicts a maximum mass $M_{\rm \textsc{tov}}$ that the NS can sustain, above which gravitational collapse leads to the formation of a black hole (BH). However, various studies showed that, in the presence of uniform or differential rotation, the NS is stable up to larger values of mass, $M_{{\rm unif}}$ and $M_{{\rm thr}}$, respectively, with $M_{\rm thr} > M_{\rm unif }$~\cite{Baumgarte:1999cq,unifrot1,unifrot2}. 
The exact values of $M_{\rm unif}$ and $M_{\rm thr}$ are not fixed, as they depend on the EoS. Several works investigated the conditions that would lead to a prompt collapse. In particular, numerous studies tried to identify EoS-independent relations to describe the threshold mass, i.e., the critical mass that leads to prompt collapse to BH, as a function of $M_{\rm \textsc{tov}}$ or some parameters of the binary, such as its mass ratio, spins, or tidal deformability~\cite{Bauswein:2013jpa,Hotokezaka:2011dh,Kolsch:2021lub,Bauswein:2020xlt,Tootle:2021umi,Kashyap:2021wzs,Bauswein:2020aag,Agathos:2019sah}.

For cold NSs, the merger remnant is usually classified into four possible results~\cite{Hotokezaka:2013iia,Sarin:2020gxb} (see also Fig.~\ref{fig:fancy_plot}). If the mass of the remnant formed after the BNS merger is larger than $M_{\rm thr}$, the remnant will collapse to a BH within the dynamical timescale of 1-2 ms. If, instead, the mass is larger than $M_{\rm unif }$ but still smaller than $M_{\rm thr}$, the remnant hypermassive NS (HMNS) can survive for tens to hundreds of milliseconds.
For remnant's mass values between $M_{\rm unif}$ and $M_{\rm \textsc{tov}}$, a supra-massive NS (SMNS) is formed, which can survive for seconds before collapsing to a BH. Finally, if the mass of the remnant is smaller than $M_{\rm \textsc{tov}}$, a stable NS is formed.

After the formation of a  HMNS or SMNS, the emission of GWs, together with angular momentum dissipation, make the remnant unstable, leading to the collapse to a BH~\cite{Hotokezaka:2013iia}.
In the case of HMNSs, also angular momentum transport processes, such as hydrodynamical effects, magnetic winds, and magnetorotational instabilities~\cite{Baumgarte:1999cq,Duez:2004nf,Shibata:2005mz,Siegel:2013nrw,Balbus:1998ja,Shibata:2017xht,Radice:2017zta}, play an important role, since they reduce the degree of differential rotation. These various processes happen on different timescales, which translates into different lifetimes $\tau_{\rm BH}$ for the remnant. Therefore, following Ref.~\cite{Hotokezaka:2011dh}, we further distinguish between long-lived HMNSs, with $\tau_{\rm BH} > 5$~ms, and short-lived HMNSs, with $\tau_{\rm BH} < 5$~ms.

Finally, also temperature effects should be taken into account, given that, after the merger, the remnant temperature can reach 30-50 Mev~\cite{Sekiguchi:2011mc,Sekiguchi:2011zd,Kaplan:2013wra}. For example, for remnants with relatively low masses, the GWs emission and the angular momentum transport and dissipation processes might not be enough to cause the collapse to BH; in this case, neutrino cooling can remove the thermal energy that still prevents the remnant from collapsing in a timescale of a few seconds~\cite{Hotokezaka:2013iia}.

Knowing the kind of remnant produced is interesting for various reasons:
\begin{itemize}
    \item The fate of a BNS system after the merger depends mainly on the combination of its total mass and EoS, because the latter determines not only $M_{\rm \textsc{tov}}$, but also $M_{\rm unif}$ and $M_{\rm thr}$. Therefore, from the collapse time, one can deduce information about the EoS. 
    \item The different possible lifetimes of the remnant are linked to the various processes that lead to its collapse to a BH. Thus, distinguishing for how long the remnant survives could help understand the physics of the postmerger phase.
    \item The lifetime of the remnant plays a crucial role in the potential electromagnetic (EM) counterparts; cf. Ref.~\cite{Margalit:2019dpi} and references therein.
    \item In the third-generation (3G) detectors era, $\mathcal{O}(10^4-10^5)$ BNS detections are expected per year~\cite{Branchesi:2023mws,Samajdar:2021egv}, therefore knowing in which cases we expect a remnant, and hence an EM counterpart, is crucial in order to point telescopes to the right sources.
    \item To build a model for the GW postmerger signal, in many cases, especially when quasi-universal relations are involved, it is necessary to know whether a postmerger signal is present or not.
\end{itemize}

In this work, we employ machine learning, and in particular the Gradient Boosted Decision Tree (GBDT)~\cite{gbdt,FRIEDMAN2002367} algorithm in \texttt{sklearn}~\cite{scikit-learn}, to build a classifier that can predict the outcome of a BNS system merger based on the system's parameters inferred from GW signals.

Refs.~\cite{Clark:2014wua,Tringali:2023ray} proposed different methods to discriminate between different postmerger scenarios, based on an analysis of the postmerger GW signal. On the other hand, machine learning techniques are increasingly employed in GW data analysis (see, e.g., Refs.~\cite{Cuoco:2020ogp,Benedetto:2023jwn,Zhao:2023tqr} for a review). For example, machine-learning methods have been proposed to search for GW signals emitted by long-lived remnants~\cite{Miller:2019jtp}, and to improve the calculation of the probability of the presence of a NS component or some remnant matter in low-latency analyses~\cite{Chatterjee:2019avs}. Moreover, recent studies have employed machine learning to construct models for the postmerger GW signal, with promising results~\cite{Pesios:2024bve, Soultanis:2024pwb}. Our goal, instead, is to predict the lifetime of the remnant based only on information that can be deduced from the inspiral GW signal. In this way, we would be able to gather information about the remnant also in the case of a BNS detection that does not include the postmerger, which is the most likely scenario with current detectors, and will also happen regularly for 3G ones, given the difficulties in measuring the high-frequency postmerger signal.

The paper is structured as follows: in Sec.~\ref{sec:methods}, we explain how the dataset is built and outline the basic features of the GBDT algorithm; in Sec.~\ref{sec:results}, we show the performance of the classifiers we built, the predictions we obtain when applying them to real events detections, and we investigate the influence of the binary parameters on the merger outcome. In Sec.~\ref{sec:uncertainty}, we discuss the uncertainty of our predictions, and we provide more details in Appendix~\ref{sec:appendix_unc}. Finally, we conclude in Sec.~\ref{sec:conclusions}.

The data and the models contained in this work are publicly available at \cite{data_release}.

\section{Methods}
\label{sec:methods}

\begin{figure*}[htb]
        \centering
        \includegraphics[width=\textwidth]{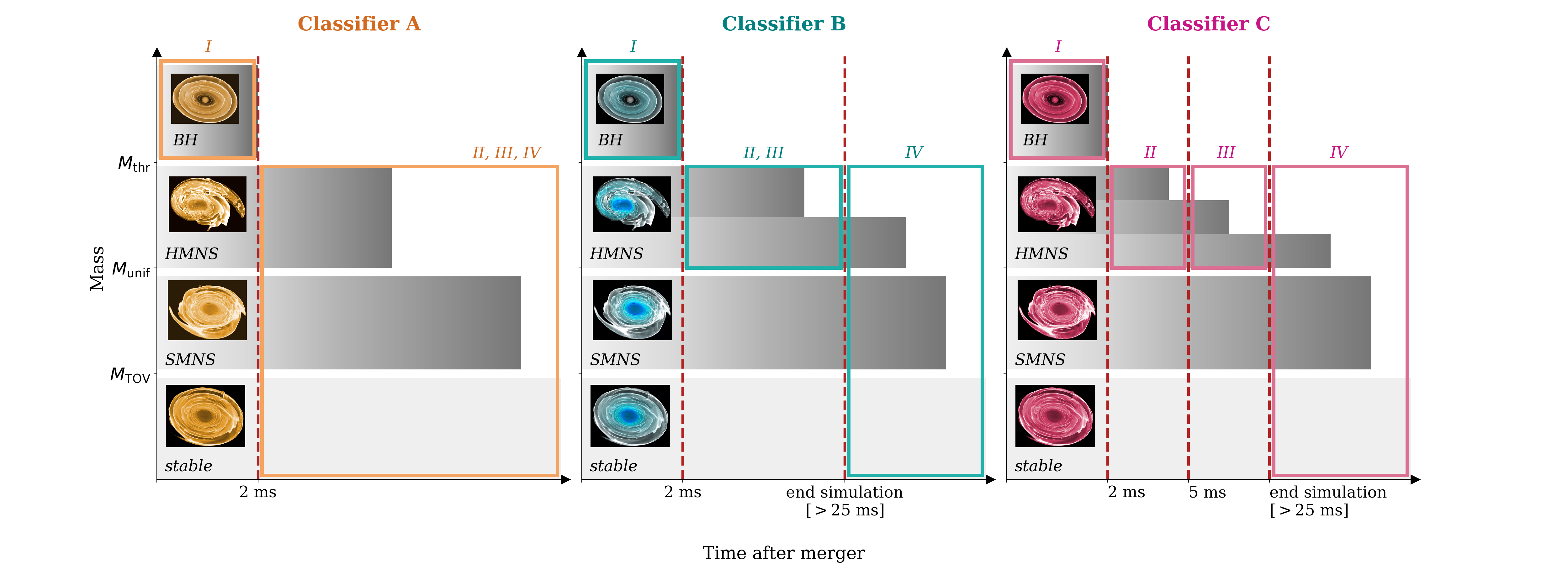}
    \caption{Schematic representation of the possible outcomes of a BNS merger depending on the system's total mass. We show the different kinds of remnants with grayscale bars that gradually progress toward black as the system evolves toward the final collapse to a BH. The red vertical dashed lines represent the thresholds on the remnant lifetime employed to distinguish the different classes for each classifier. The colored boxes (orange for Classifier A, teal for Classifier B, and pink for Classifier C) show how the various remnants are grouped in the different classes employed for each classifier.}
    \label{fig:fancy_plot}
\end{figure*} 

In this section, we describe how we build the dataset, with a focus on the procedure employed to split the data points into training and validation sets. We briefly illustrate how the GBDT algorithm works and explain how we evaluate the performance of our models.

\subsection{Dataset}
\label{ssec:dataset}

To build our dataset, we employ numerical-relativity (NR) simulations data from the CoRe~\cite{Dietrich:2018phi,core_webpage} and SACRA~\cite{Kiuchi:2019kzt} databases, together with simulations from Refs.~\cite{more_nr, Schianchi:2024vvi}. The simulations considered span a wide range of values for the binary system's parameters and employ different EoSs, also including phase transitions. When the remnant collapse time is not directly provided, we compute it from the GW data, as the difference between the merger time and the time when the GW signal goes to zero (see the bottom panel of Fig.~\ref{fig:class_distr}). If the GW signal does not go to zero, we register the remnant lifetime as ``greater'' than the simulation time after the merger, computed as the difference between the merger time and the total simulation time, after removing the last 2~ms to take into account the effects of data tapering. Regarding the data in Refs.~\cite{more_nr, Schianchi:2024vvi}, for the configurations that do not collapse within the simulation time, we consider the remnant lifetime to be $> 25 \, {\rm ms}$, where $25 \, {\rm ms}$ is a conservative estimate of the duration of these simulations after the merger.
If the system total mass is lower than $M_{\rm \textsc{tov}}$ for the specific EoS employed in the simulation, we classify the remnant as ``stable''.
For each system, we consider data from the simulation with the highest resolution and the largest extraction radius available\footnote{In the case of a few configurations that were included both in the CoRe database and in Refs.~\cite{more_nr, Schianchi:2024vvi}, we kept the collapse time value provided in Refs~\cite{more_nr, Schianchi:2024vvi}, which is computed directly from the NR data describing the system evolution. More details about the selected simulations can be found at \cite{data_release}.}. We discard eccentric configurations, and, among the simulations in Ref.~\cite{Schianchi:2024vvi}, we do not include the ones for misaligned-spin configurations. In the end, our dataset includes 398 systems.

The system parameters employed as features for the classifier are the total mass $M_{\rm tot}$, the mass ratio $q = m_1/m_2$, with $m_1 \ge m_2$, the mass-weighted tidal deformability $\tilde{\Lambda}$, and the effective spin $\chi_{\rm eff}$. The latter two are defined, respectively, as~\cite{Wade:2014vqa,Favata:2013rwa,Santamaria:2010yb}
\begin{equation}
	\tilde{\Lambda} = \frac{16}{13} \frac{(m_1 + 12 m_2)m_1^4 \Lambda_1 + (m_2 + 12 m_1) m_2^4 \Lambda_2}{(m_1 + m_2)^5},
\end{equation}
where $\Lambda_{1,2}$ are the components tidal deformabilities,  and
\begin{equation}
	\chi_{\rm eff} = \frac{\left(m_1 \chi_{1, \parallel} + m_2 \chi_{2, \parallel}\right)}{m_1 + m_2},
\end{equation}
with $\chi_{1,\parallel}, \chi_{2,\parallel}$ being the spin components parallel to the angular momentum.

Following Refs.~\cite{more_nr,Hotokezaka:2011dh}, we classify the merger remnant in four possible cases, based on its collapse time $\tau_{\rm BH}$:
\begin{enumerate}[I.]
    \item $\tau_{\rm BH} < 2 \, {\rm ms}$: prompt collapse,
    \item $2 \, \rm {ms} < \tau_{\rm BH} < 5 \, {\rm ms}$: short-lived HMNS,
    \item $\tau_{\rm BH} > 5 \, {\rm ms}$: long-lived HMNS,
    \item no collapse within simulation time (that can happen for HMNSs, SMNSs, or stable NS remnants).
\end{enumerate}

We build three separate classifiers to distinguish between different scenarios:
\begin{enumerate}[A.]
    \item \emph{Classifier A} discriminates between prompt collapse to BH (case I) \textit{vs.} formation of a NS remnant (case II, III, IV); see Sec.~\ref{ssec:cl_a}.
    \item \emph{Classifier B} discriminates between prompt collapse to BH (case I) \textit{vs.} formation of a HMNS (case II, III) \textit{vs.} no collapse within simulation time (case IV); see Sec.~\ref{ssec:cl_b}.
    \item \emph{Classifier C} discriminates between all four different scenarios listed above; see Sec.~\ref{ssec:cl_c}.
\end{enumerate}

Figure~\ref{fig:fancy_plot} shows the possible outcomes of a BNS merger and what the different classifiers try to distinguish.

\begin{figure*}[htb]
        \centering
        \includegraphics[width=\textwidth]{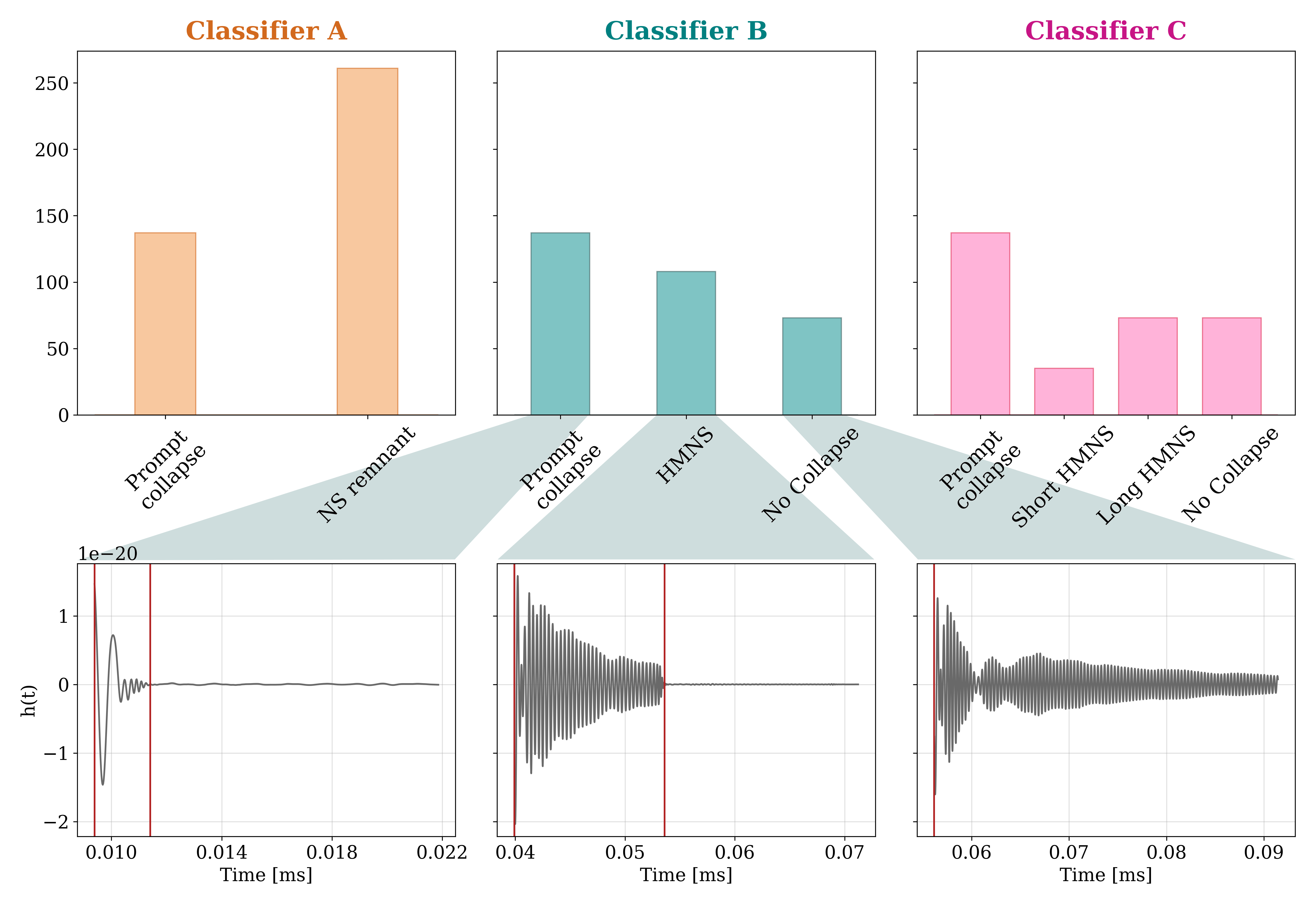}
    \caption{Top panel: distribution of points in the dataset in the different classes for each classifier. For Classifier A the class corresponding to formation of a NS remnant includes clearly many more entries than the prompt collapse one, while for Classifier C the short-lived HMNS class is the one with fewer points. Bottom panel: example of postmerger gravitational waveform for simulations in the CoRe database, one for each class of Classifier B, i.e., from left to right, prompt collapse to BH, formation of a HMNS that collapses before the end of the simulation, and formation of a NS remnant that does not collapse within simulation time. The red, vertical lines correspond to the merger time and to the time where the GW signals goes to zero. 
    Note that for the last waveform on the right, the collapse time is not shown since it does not occur within the simulation.}
    \label{fig:class_distr}
\end{figure*}

\subsection{Data splitting}
\label{ssec:data_split}

In order to build our classifier, we split the data in $90 \%$ for the training set and $10 \%$ for the validation set. However, we must take into account the fact that our labels distribution is unbalanced (see Fig.~\ref{fig:class_distr}). To avoid possible biases, meaning that the model is trained to predict mostly systems belonging to only one class, we split the dataset with the \texttt{sklearn} tool `\texttt{train\_test\_split}' and the \textit{`stratify'} option, specifically designed to avoid unbalanced data in the training-validation splits~\cite{split_func}.

Moreover, the performance of the classifier is strongly influenced by the random seed utilized in the training-validation splitting. This is due to the fact that some data points are more difficult to classify, for various reasons (for example, when $\tau_{\rm BH}$ is close to the threshold values that determine the class, or when some configurations have the same parameters but different spin values and can, therefore, result in different outcomes). This effect can be mitigated by ensuring that at least some of the points with ``problematic'' features are included in the training set, in order to improve their modeling. On the other hand, we also want to include some of them in the validation set, to make sure that the evaluation of the classifier performance is not carried out only on ``easy'' points and thus biased. 
To accomplish both these goals, we create 1000 distinct training-validation splittings, using a different seed for each one. For each splitting, we train and validate a GBDT model with a fixed random state and with the default parameters provided for the \texttt{GradientBoostingClassifier} class in \texttt{sklearn}\footnote{Optimizing the hyperparameters for each model would decrease the number of misclassified points, but at the expense of a much larger computational cost. Hence, we keep the default parameters, with the resulting misclassifications representing a conservative scenario. On the other hand, the random state, which controls the randomness present in various steps of the algorithm, is not fixed by default in the \texttt{sklearn} model. We specify it not only to ensure reproducibility, but also to avoid possible additional sources of uncertainty, since at this stage we want to focus on the variations induced by the different training-validation splittings. We arbitrarily set \texttt{random\_state}$=42$; for completeness, we checked the effect of different values for the GBDT model random state and found that results are not significantly affected.}, and save the misclassified points as ``bad'' points. We further divide the final batch of ``bad'' points in three groups based on the number of their misclassification occurrences: less than 10 times, between 10 and 90 times, and more than 90 times, with the cutoff values chosen arbitrary. The final training (validation) set is built by taking $90 \%$ ($10\%$) of the points that were never misclassified, together with $90\%$ ($10 \%$) of the points in each one of the ``bad'' points groups. 
With this procedure to include ``bad'' points in both the training and validation sets, we consider again different splittings created with different random seeds, and, to train and test the classifier, we choose among these one of the training-validation splittings yielding the highest validation accuracy.
This approach ensures that the model is trained to recognize also the difficult points, and that the final evaluation for the model performance on the validation set is not biased. 

Table~\ref{tab:ranges} reports the ranges for parameters in both the training and complete datasets for all three classifiers. We highlight that the distribution of points in our dataset is not uniform in these ranges, since it depends on the available NR simulations.

\color{red}
\begingroup
\renewcommand*{\arraystretch}{2}
\begin{table*}[htb]
\begin{tabular}{c cc cc cc}
             & \multicolumn{2}{c}{Classifier A} & \multicolumn{2}{c}{Classifier B} & \multicolumn{2}{c}{Classifier C}\\
             \hline
             \hline
             & Training          & Total            & Training     & Total      & Training     & Total  \\
             \hline
$M_{\rm tot}$  & [2.4, 3.4] $M_{\odot}$ &[2.4,3.4] $M_{\odot}$ &[2.4,3.4] $M_{\odot}$  & [2.4, 3.4] $M_{\odot}$  &[2.4, 3.4] $M_{\odot}$  &[2.4,3.4] $M_{\odot}$    \\
$q$   & [1.0, 2.1]  &[1.0, 2.1] &[1.0, 2.0] &[1.0,2.0] &[1.0, 2.0] &[1.0,2.0] \\
$\tilde{\Lambda}$   &[89, 3520]  &[89, 6255] &[89, 3520]  &[89, 6255] &[89, 6255] &[89, 6255]\\
$\chi_{\rm eff}$   &[-0.267, 0.409] &[-0.267, 0.409] &[-0.267, 0.272] &[-0.267, 0.272] &[-0.267, 0.267] &[-0.267, 0.272]
\end{tabular}
\caption{Ranges of parameters covered by the training and total dataset for the different classifiers. As will be discussed in Sec.~\ref{ssec:cl_b}, the dataset employed for Classifier B and Classifier C is different from the dataset for Classifier A, since it does not include simulations that do not collapse within simulation time, for which the simulation time after the merger is $t_{\rm sim} < 25$~ms; consequently, the ranges in the ``Total'' columns are slightly different.}
\label{tab:ranges}
\end{table*}
\endgroup
\color{black}

\subsection{Gradient Boosted Decision Trees Classifier}

Decision trees are a non-parametric supervised learning technique, whose output is based on a series of hierarchical decision rules inferred from the data~\cite{murthy1998automatic}. In ensemble methods, such as random forests or boosting methods~\cite{breiman2001random,Freund1990BoostingAW,friedman2001greedy}, multiple decision trees are combined, in order to improve the robustness of the predictions.
In \emph{Gradient Boosted Decision Trees}, the prediction for an input $\mathbf{x}_i$ is given by
\begin{equation}
    y_i = F_\textsc{m}(\mathbf{x}_i) = \sum_{m=1}^M h_m(\mathbf{x}_i),
\end{equation}
where the \emph{weak learners} $h_m$ are single decision trees, $M$ represents the total number of weak learners, and the ensemble model $F_\textsc{m}$ is also called the \emph{strong model}\footnote{Decision trees predictions are continuous values, hence, in classification problems, the final prediction, i.e., the discrete value corresponding to a class, needs to be computed from the continuous $F_\textsc{m}(\mathbf{x}_i)$ with a function depending on the loss function employed. In the case of a log-loss error, a sigmoid and softmax functions are used for binary and multiclass classification problems, respectively~\cite{friedman2001greedy}.}.
The gradient boosting model is built through an iterative approach~\cite{friedman2001greedy}, with the goal to minimize the expectation value of the loss function $L$, which in our case is a log-loss, on the dataset $\left\{ \mathbf{x}_i, y_i\right\}_{i=1}^{N}$. At each step $t$, the strong model is updated with a new weak learner
\begin{equation}
    F_{t} = F_{t-1} - \epsilon \, h_{t},
\end{equation}
where the \emph{shrinkage ratio} (or \emph{learning rate} in \texttt{sklearn}) $\epsilon\in[0,1]$ is a parameter introduced to reduce overfitting. The weak learner is chosen from a family $H$ of decision trees in order to minimize the average of the loss function on the dataset of the strong learner at the previous step
\begin{equation}
    h_{t} = \arg \min_{h\in H} \sum_{i=1}^N L(y_i, F_{t-1}(\mathbf{x}_i) + h(\mathbf{x}_i)).
\end{equation}
Applying a steepest-descent step to this minimization problem, and Taylor-expanding the loss function, one finds that the local maximum-descent direction of the loss function is given by its negative gradient. Hence, at each step, the weak learner is found as
\begin{equation}
    h_t \sim \arg\min_{h\in H} \sum_{i=1}^N h(\mathbf{x}_i) \left[ \frac{\partial L(y_i, F(\mathbf{x}_i))}{\partial F(\mathbf{x}_i)} \right]_{F=F_{t-1}}.
\end{equation}

The performance of GBDT models can be improved by tuning their hyperparameters (see, e.g., Ref.~\cite{hastie2009elements}). Regarding the \emph{shrinkage ratio} $\epsilon$, although larger values imply that the model learns faster, smaller values help avoid overfitting; typically, we employ $\epsilon \sim 0.1 $. Another useful parameter is the total number $M$ of weak learners comprising the model (dubbed \texttt{n\_estimators} in \texttt{sklearn}): to avoid overly complex models, an adequate value of $M$ can be estimated via \emph{early stopping}~\cite{Morgan1989GeneralizationAP,Reed1993PruningAS,Zhang2005BoostingWE}; for our classifiers, typical values of $M$ are around 100. Finally, the complexity and size of each weak learner is controlled by the \texttt{max\_depth} parameter, which limits the number of nodes in each tree. In GBDT, the component trees are usually quite shallow, and we use \texttt{max\_depth} $<5$.

\subsection{Performance of the classifier}

Different metrics can be employed to assess the performance of a classifier. The most common one is probably the \emph{accuracy}, computed as the percentage of the correctly predicted labels
\begin{equation}
    \alpha (y,\hat{y}) = \frac{1}{N} \sum_{i=0}^{N - 1} \mathbf{1}(\hat{y}_i = y_i),
\end{equation}
where $\hat{y}_i$ is the predicted label for the $i$-th sample, $y_i$ is its true label, $N$ is the total number of samples, and $\mathbf{1}$ is the indicator function.
The accuracy, however, does not take into account the distribution of correct and wrong predictions across the different classes, and the information it provides can be misleading in the case of highly imbalanced datasets.

Let us consider the case of a binary classification, where usually the predictions are labeled as ``positive'' (or 1, which in our case could be the presence of a NS remnant) and ``negative'' (or 0, in our case the prompt collapse to BH). The confusion matrix obtained with our model on a given set of data will include four different kinds of entries: true positives (TP), the data points correctly classified as positive, false positives (FP), data points whose true label is negative but are misclassified as positive, true negatives (TN), points correctly classified as negative, and false negatives (FN), positives misclassified as negative. In the binary case, the accuracy is
\begin{equation}
    \alpha = \frac{TP + TN}{ TP + FP + TN + FN}.
\end{equation}
If the dataset, for example, comprises 95\% points with label ``positive'' and only 5\% ``negative'', also if all the negative points are misclassified as positive, we would still retrieve a very large value of accuracy. 

\begin{figure}[htb]
    \begin{minipage}[]{.48\textwidth}
        \centering
        \includegraphics[width=\textwidth]{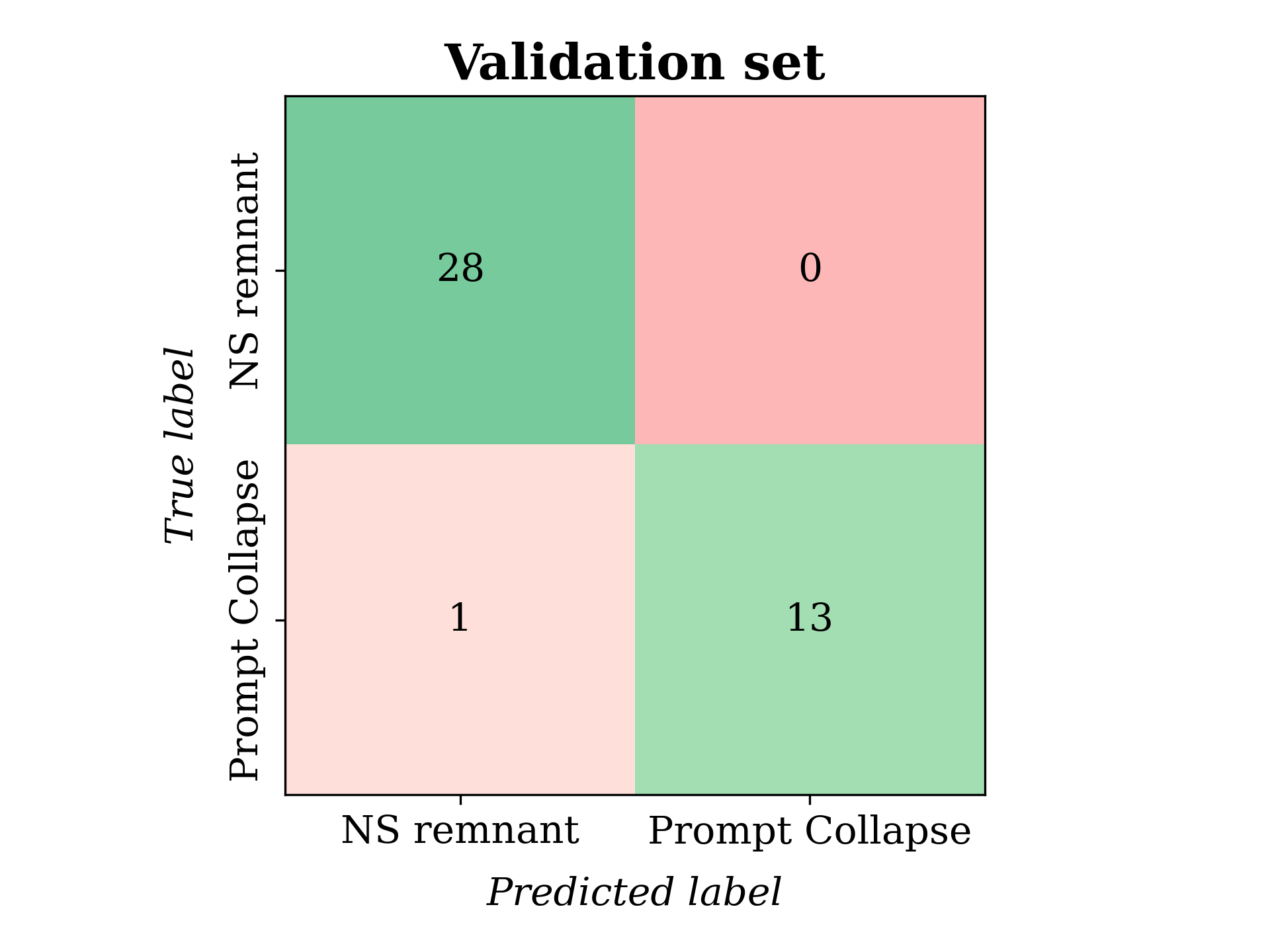}
    \end{minipage}
    \hfill
    \begin{minipage}[]{.48\textwidth}
        \centering
        \includegraphics[width=\textwidth]{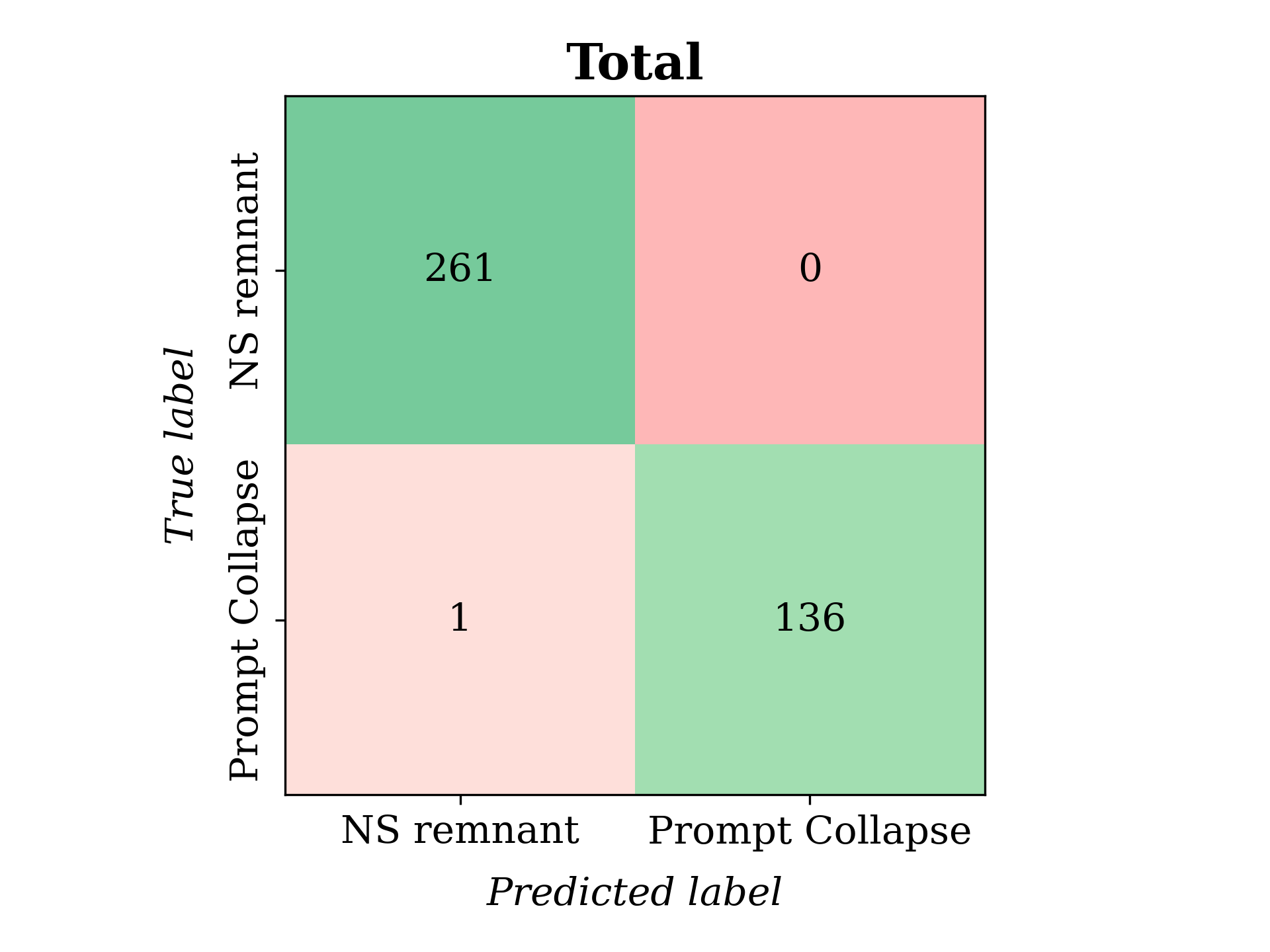}
    \end{minipage}  
    \caption{Confusion matrix for Classifier A, computed on the validation set (top panel) and on the complete dataset (bottom panel). The green color on the diagonal line highlights the entries that are correctly classified, while the red corresponds to the matrix elements with potential misclassifications.}
    \label{fig:confmatr_prompt}
\end{figure}

A more robust metric should take into account all possible entries of the confusion matrix. The \emph{Matthews correlation coefficient}~\cite{MATTHEWS1975442,mcc_overview}, for a binary classification problem, is defined as
\small
\begin{equation}
    MCC = \frac{(TP \times TN) - (FP \times FN)}{\left[(TP + FP)(TP+FN)(TN+FP)(TN+FN)\right]^{1/2}}.
\end{equation}
\normalsize
MCC values range between $-1$ and $1$, indicating the worst and best performance, respectively, while $MCC=0$ corresponds to a random prediction. The Matthews correlation coefficient can be generalized to multiclass classification tasks as
 \begin{equation}
     MCC = \frac{c \times s - \sum_k p_k \times t_k}{(s^2 - \sum_k p_k^2) \times (s^2 - \sum_k t_k^2)},
 \end{equation}
with the $k$ index representing the different classes. In the definition above, $t_k$ is the number of times a class $k$ truly occurred, $p_k$ the number of times a class $k$ was predicted, $c$ is the total number of samples correctly predicted, and $s$ the total number of samples in the evaluated dataset. In the multiclass scenario, the maximum value for $MCC$ remains $+1$, but the minimum value lies between $-1$ and $0$, depending on the number of classes and the distribution of labels.

\begingroup
\renewcommand*{\arraystretch}{2}
\begin{table}[htb]
\begin{tabular}{p{8em} cc cc}
             & \multicolumn{2}{c}{Validation set} & \multicolumn{2}{c}{Total} \\
             \hline
             \hline
             & $\alpha$          & MCC            & $\alpha$      & MCC       \\
             \hline
Classifier A  & 97.6\%            & 0.946          & 99.8\%        & 0.994     \\
Classifier B  & 94.1\%            & 0.909          & 99.1\%        & 0.985     \\
Classifier C  & 88.2\%            & 0.831          & 97.5\%        & 0.964         
\end{tabular}
\caption{Accuracy $\alpha$ and Matthews Correlation Coefficient ($MCC$) for the different classifiers, computed both on the validation set and on the total dataset, including the training points.}
\label{tab:accuracies}
\end{table}
\endgroup

\section{Results}
\label{sec:results}

In this section, we discuss the three classifiers, in terms of both their accuracy and predictions on the two real events GW170817 and GW190425. For Classifier A, we also investigate the influence of the binary parameters in determining whether the system will promptly collapse to a BH or not.

\begin{figure*}[htb]
        \centering
        \includegraphics[width=0.8\linewidth]{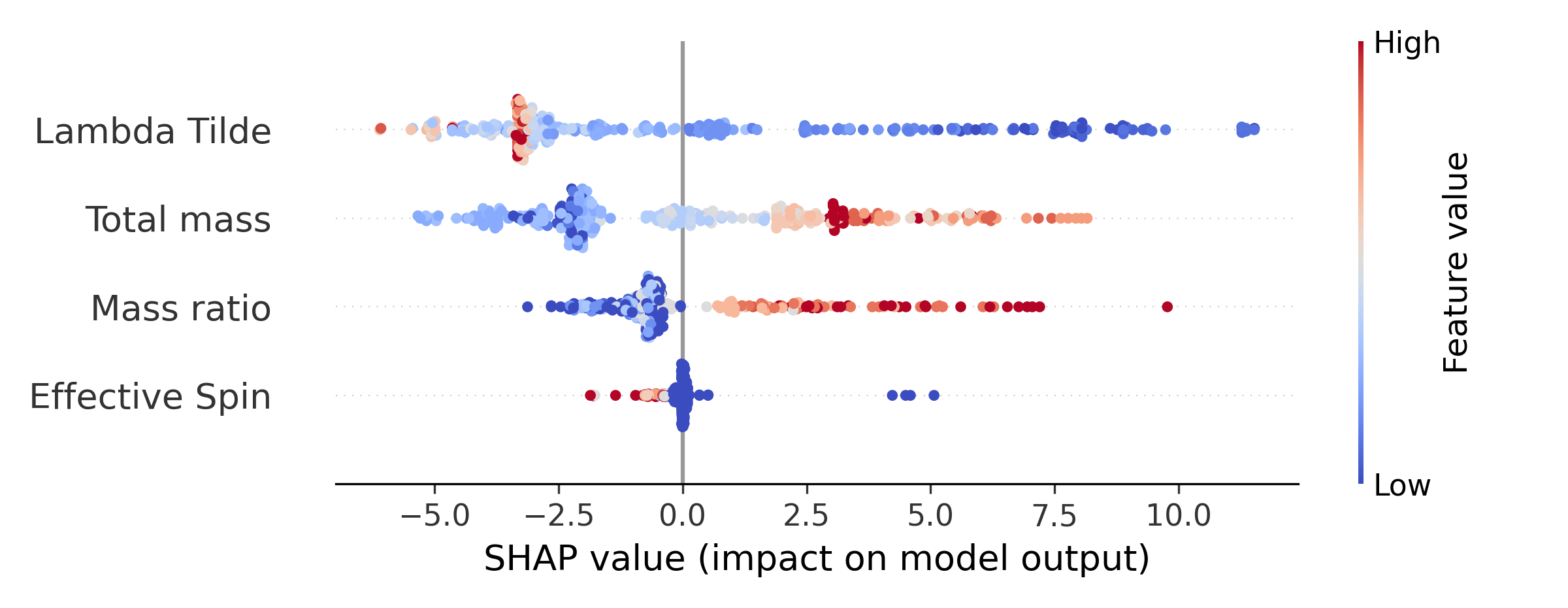}
    \caption{Summary plot for the computed SHAP values, where each dot in the plot corresponds to a point in the training set. The features are ordered from the top based on their mean SHAP absolute value, with the color representing the feature values. Larger, positive (smaller, negative) SHAP values indicate that the model predicts larger (smaller) log odds values, which in our case translate into a larger probability for the prompt collapse (formation of a NS remnant) scenario.}
    \label{fig:shap_summary}
\end{figure*}

\subsection{Classifier A}
\label{ssec:cl_a}
As a first step, we build a classifier to distinguish between prompt collapse and the formation of a NS remnant. We divide the 398 points into a training and a validation set as explained in Sec.~\ref{ssec:data_split}, and then we optimize the hyperparameters of the model using \texttt{GridSearchCV}~\cite{grid_search}, manual tuning, and \emph{early stopping}. 

Figure~\ref{fig:confmatr_prompt} shows the confusion matrix, i.e., a summary of the correct and wrong classifications of the model, obtained on the validation set (top panel), and, for completeness, on the total dataset, including the training points (bottom panel). Overall, we achieve very high values in the performance indicators: $\alpha = 97.6\%$ and $MCC = 0.946$ on the validation set, $ \alpha = 99.8 \%$ and $MCC = 0.994$ on the total one. These values are also reported in Table~\ref{tab:accuracies}, together with the ones obtained for the other classifiers.

The \texttt{sklearn} GBDT algorithm provides also the features importances for the model, computed as the Gini importance, and we find that the parameter that affects the classification the most is $\tilde{\Lambda}$ ($\sim 0.60$), followed by $M_{\rm tot}$ ($\sim 0.20$), $q$ ($\sim 0.14 $), and finally $\chi_{\rm eff}$ ($\sim 0.04$). Given the possible inconsistency among values of the features' importance computed with different methods (e.g., gain, split count, or permutation method)~\cite{lundberg2018consistent}, a new measure of the contribution of each feature to a model prediction has been proposed, the SHAP (SHapley Additive exPlanations) values~\cite{lundberg2017unified}. The SHAP method derives from game theory~\cite{trumbelj2014ExplainingPM} and belongs to the Explainable Artificial Intelligence research field~\cite{arrieta2020explainable}, which investigates possible methods to explain models obtained with artificial intelligence tools that would otherwise be considered ``black-boxes''. The SHAP technique assigns locally, i.e., for each data point, an importance to each feature based on its impact on the model's prediction for that point. For Classifier A, we compute SHAP values for the points in the training set with the \texttt{shap} Python package~\cite{lundberg2017unified,lundberg2018consistent,shap_docs}.\footnote{For the moment, the computation of SHAP values for GBDT models is supported only for binary classifiers.} The SHAP values are related to the log odds predicted by the model: more positive (negative) SHAP values ``push'' the model prediction to larger positive (negative) values, which, for Classifier A, translate to a larger probability for the prompt collapse to BH (production of a NS remnant) scenario. Therefore, to evaluate the importance of a feature, one needs to look at the absolute value of its SHAP measure. Computing the mean absolute SHAP values for each feature, we find again that the parameter with the largest impact on the model prediction is $\tilde{\Lambda}$, followed by $M_{\rm tot}$, $q$, and $\chi_{\rm eff}$. However, if we look at the maximum of the SHAP absolute values, the second most important feature is the mass ratio $q$. This means that, on average, $M_{\rm tot}$ has a stronger impact than $q$, but, for the single data points, $q$ can actually affect the prediction more than $M_{\rm tot}$. Figure~\ref{fig:shap_summary} shows the summary plot for the computed SHAP values. In addition to the order of the features based on their importance, which we already discussed, we see that, as expected, higher values for $M_{\rm tot}$ are more likely to push the prediction toward prompt collapse and the same holds for large $q$ values. The effective spin $\chi_{\rm eff}$ is clearly the feature with the more modest impact, and indeed most of the points are clustered around very small absolute values of SHAP. Although, on average, the SHAP values are small, for a few points, the impact of $\chi_{\rm eff}$ is non-negligible. We notice that negative spin values tend to influence the prediction toward a prompt collapse, while positive ones toward the production of an NS remnant; we will discuss the spin effect in more detail later.

Although one might expect the total mass to have a more significant impact, what determines the fate of the BNS after the merger is the value of the total mass compared to the various thresholds for the maximum supported mass, $M_{\textsc{tov}}$, $M_{\rm unif}$, and $M_{\rm thr}$, which depend on the EoS. Information about $M_{\textsc{tov}}$ would be essential to exactly determine the outcome of the merger, but it cannot be inferred from the inspiral GW signal. Instead, in our analysis, all the information about the EoS is encoded in the $\tilde{\Lambda}$ parameter. 

We also note that only a limited number of NR simulations for spinning systems are available, thus, the very weak contribution of spin to the overall classification might not necessarily be interpreted as a weak influence of this parameter on the remnant fate, but rather as the need for more spinning simulations to better understand the role of such parameter; cf. works in Ref.~\cite{Tootle:2021umi, Schianchi:2024vvi}.

\begingroup
\renewcommand*{\arraystretch}{2}
\begin{table*}[]
\begin{tabular}{l cc cc cc ccc cc cccc}
                         \hline
                         \hline
                         & & & \multicolumn{2}{c}{Classifier A} & & & \multicolumn{3}{c}{Classifier B} & & & \multicolumn{4}{c}{Classifier C}\\
                         \hline
                         & & & $p_{\textsc{pcbh}}$ & $p_\textsc{rns}$ &  &  & $p_{\textsc{pcbh}}$  & $p_{\textsc{hmns}}$ & $p_{\textsc{nc}}$ &  &  & $p_{\textsc{pcbh}}$ & $p_{\textsc{short}}$ & $p_{\textsc{long}}$ & $p_{\textsc{nc}}$ \\
                          \hline
                          \hline
GW170817                  & & &  39.8\% & 60.2\%  & & & 39.7\% & 57.5\% & 2.8\% & & & 41.7\% & 15.6\% & 39.8\% & 3.7\% \\
GW170817+EoS            & & &  9.0\%  & 91.0\%  & & & 8.8\% & 90.5\% & 0.7\% & & & 11.6\% & 38.7\% & 49.2\% & 0.5\%\\\
GW170817+EoS+KN       & & & 0.9\% & 99.1\%  & & & 0.9\% & 98.9\% & 0.2\% & & & 1.3\% & 42.8\% & 55.8\% & 0.2\%\\
GW170817+EoS+KN+GRB    & & & 0.1\% & 99.9\% & & & 0.2\% & 99.5\% & 0.3\% & & & 0.5\% & 50.8\% & 48.6\% & 0.1\% \\
                          \hline
GW190425                  & & & 59.5\%  & 40.5\%  & & & 66.2\% & 7.4\% & 26.4\% & & &  71.9\% & 0.3\% & 11.1\% & 15.7\% \\
GW190425+EoS            & & & 98.2\%  & 1.8\%  & & & 98.6\% & 0.2\% & 1.1\% & & & 97.3\% & 0.0\% & 0.1\% & 2.5\% \\
\end{tabular}
\caption{Probabilities for the different merger outcomes for the three classifiers. We remind here that Classifier A discriminates between prompt collapse to BH ($p_{\rm \textsc{pcbh}}$) and formation of a NS remnant ($p_{\rm \textsc{rns}}$); Classifier B distinguishes between prompt collapse ($p_{\rm \textsc{pcbh}}$), formation of a HMNS that collapses during the simulation ($p_{\rm \textsc{hmns}}$), and formation of a NS remnant (which could be a HMNS, SMNS or stable NS) that does not collapse within simulation time ($p_{\rm \textsc{nc}}$). Finally, Classifier C includes four classes: prompt collapse to BH ($p_{\rm \textsc{pcbh}}$), production of a HMNS that collapses during the simulation, with $2 \, {\rm ms} < \tau_{\rm BH} < 5 \, {\rm ms}$ ($p_{\rm \textsc{short}}$) or $\tau_{\rm BH} > 5 \, {\rm ms}$ ($p_{\rm \textsc{long}}$), and formation of a remnant that does not collapse within the simulation time ($p_{\rm \textsc{nc}}$). }
\label{tab:realev}
\end{table*}
\endgroup

\subsubsection{Real events: GW170817 and GW190425}

In the following, we employ our classifier to predict the merger outcome of GW170817 and GW190425, based on the sample values for the parameters, as inferred from the analysis of the measured GW signal. All the predictions regarding these two real events, including results for the other classifiers discussed in the upcoming sections, are summarized in Table~\ref{tab:realev}.

We apply our Classifier A model to the GWTC-1 posterior samples of GW170817~\cite{gw170817_samp} obtained with the \texttt{IMRPhenomPv2\_NRTidal} waveform~\cite{Hannam:2013oca,Dietrich:2017aum,Dietrich:2018uni} and the low-spin prior (see Ref.~\cite{LIGOScientific:2018hze} for details). We feed the values of $M_{\rm tot}$, $\tilde{\Lambda}$, $q$, and $\chi_{\rm eff}$ of each sample to the classifier and compute the probability of the various scenarios as
\begin{equation}
    p_{i} = \frac{n_i}{\sum_i n_i},
\end{equation}
with $n_i$ being the number of samples predicting the scenario $i$. We find that the probability of prompt collapse to BH (PCBH) and formation of a NS remnant (RNS) are $p_{\rm \textsc{pcbh}} \sim 40 \%$ and $p_{\rm \textsc{rns}} \sim 60\%$, respectively.

However, this result is derived from the GW data alone, and more stringent constraints are obtained if we include additional information about the EoS. In Ref.~\cite{Koehn:2024set}, the GW170817 and GW190425 data are reanalyzed with a prior on the EoS that takes into account, among others, constraints derived from chiral effective field theory, perturbative quantum chromodynamics, and the minimum values for $M_{\rm \textsc{tov}}$ inferred from heavy radio pulsars. If we consider the posteriors for GW170817 produced in this work with such additional constraints (GW170817+EoS), Classifier A predicts a much larger probability for the formation of a remnant, with $p_{\rm \textsc{pcbh}} \sim 9\%$ and $p_{\rm \textsc{rns}} \sim 91 \%$. The results obtained from the two sets of samples are consistent with the findings in Ref.~\cite{Agathos:2019sah}, where, with two different methods, the probability of prompt collapse to BH for GW170817 is predicted to be $\sim 50-70 \%$, but below $\sim 10\%$ when mass information from the heavy pulsars is included.

Similarly, analyzing the GWTC-2.1 samples for GW190425~\cite{gwtc21_open}, Classifier A predicts $p_{\rm \textsc{pcbh}} \sim 60 \%$ and $p_{\rm \textsc{rns}} \sim 40\%$. However, if we consider the samples in Ref.~\cite{Koehn:2024set}, produced including the additional EoS constraints (GW190425+EoS), the classifier yields a prompt-collapse probability $p_{\rm \textsc{pcbh}} \sim 98\%$, consistent with expectations of prompt-collapse being the most likely outcome for such high-mass binaries.

Additionally, Ref.~\cite{Koehn:2024set} reanalyzed GW170817 incorporating also information from the EM counterpart~\cite{LIGOScientific:2017ync}, the kilonova AT2017gfo~\cite{LIGOScientific:2017pwl} (GW170817+EoS+KN) and the short gamma-ray burst GRB170817A~\cite{LIGOScientific:2017zic} (GW170817+EoS+KN+GRB). For the posteriors produced including EM data, the predicted probability of formation of a NS remnant increases further, up to $99.9\%$ (see Table~\ref{tab:realev}).

Incorporating additional information in the analysis of GW170817 and GW190425 significantly improves the prediction accuracy mostly because the posterior probability densities obtained from GW data only are quite wide, especially for $\tilde{\Lambda}$. This translates into the presence of samples with combinations of masses and $\tilde{\Lambda}$ that are not compatible with most physical EoSs (for example, large values of $\tilde{\Lambda}$ for the quite large mass values in GW190425). Consequently, including information about the EoS places tighter constraints on the $\tilde{\Lambda}$ posteriors, producing more precise predictions. For future-generation detectors, we expect much tighter constraints on the parameters from the GW data alone (see, e.g., Ref.~\cite{Puecher:2023twf}), and therefore precise predictions also without additional information.

The inclusion of data from the kilonova and short gamma-ray burst considers information from the postmerger phase, but (i) it does not require the detection of a postmerger GW signal, and (ii) it shows an additional improvement of the analysis, it is not a requirement for the classifier usage. A potential independent detection of the postmerger EM counterpart could provide an interesting comparison with our classifiers' prediction from GW data. For example, if the classifier predicts the formation of a remnant, but the EM measurements are indicative of a prompt collapse, this might point to a phase transition happening with the increased density in the postmerger phase. Nevertheless, to investigate this kind of effect, an extended study of all possible systematics would be needed, which requires a larger number of NR simulations that are not available at the moment.

\begin{figure}[htb]
        \centering
        \includegraphics[width=\linewidth]{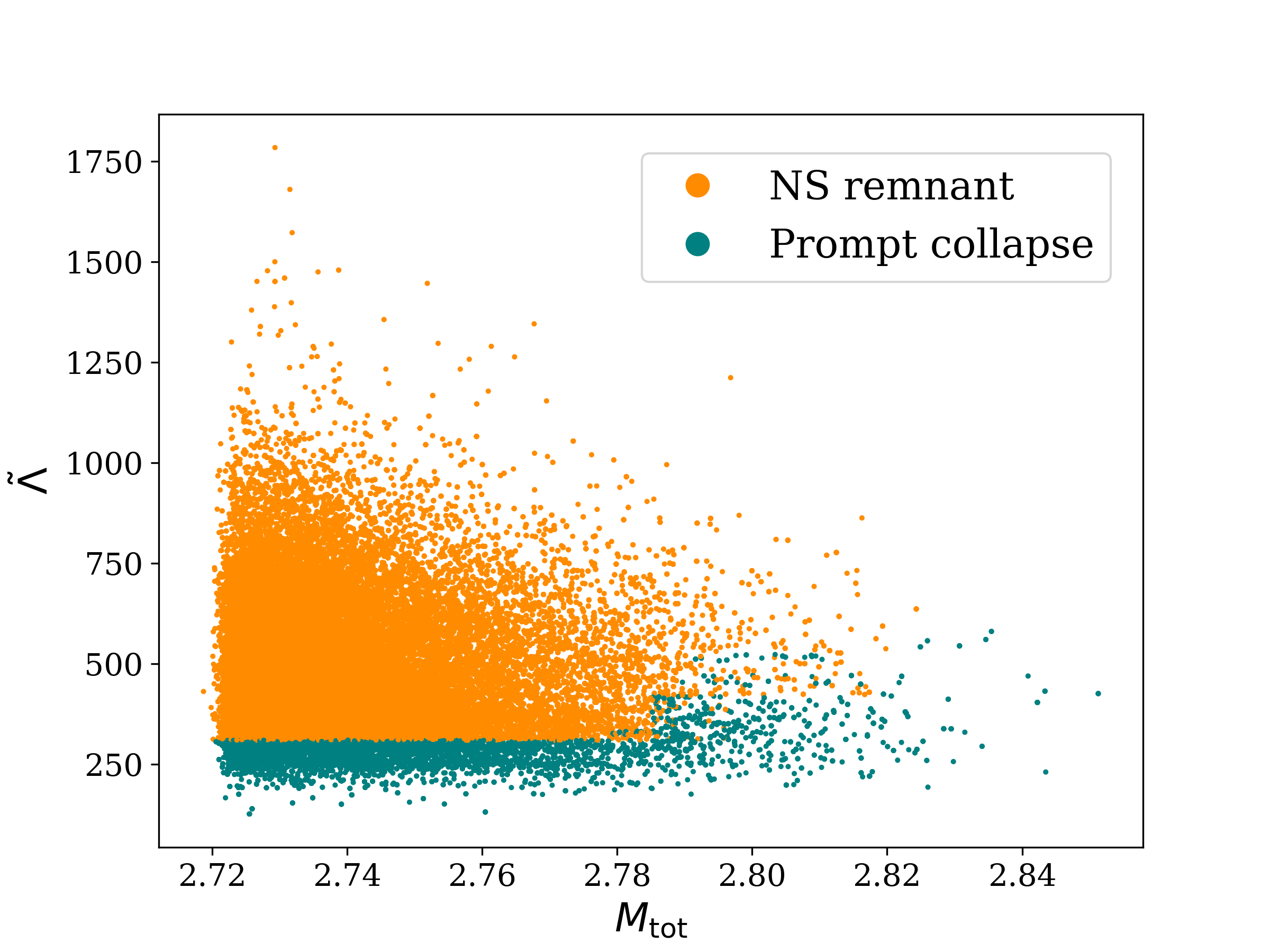}
    \caption{Distribution of $M_{\rm tot}$ and $\tilde{\Lambda}$ sample values produced in Ref.~\cite{Koehn:2024set} for GW170817+EoS, where the different colors indicate that Classifier A predicted for that sample a prompt collapse to BH (teal) or formation of a NS remnant (orange). Note that we show only the $\tilde{\Lambda}$ and $M_{\rm tot}$ values of the samples since they are the most relevant ones for the classifier, but the input features include also the sample $q$ and $\chi_{\rm eff}$ values, which are not shown here for simplicity.}
    \label{fig:mass_lambda_gw170817}
\end{figure}

\begin{figure*}[htb]
        \includegraphics[width=0.32\textwidth]{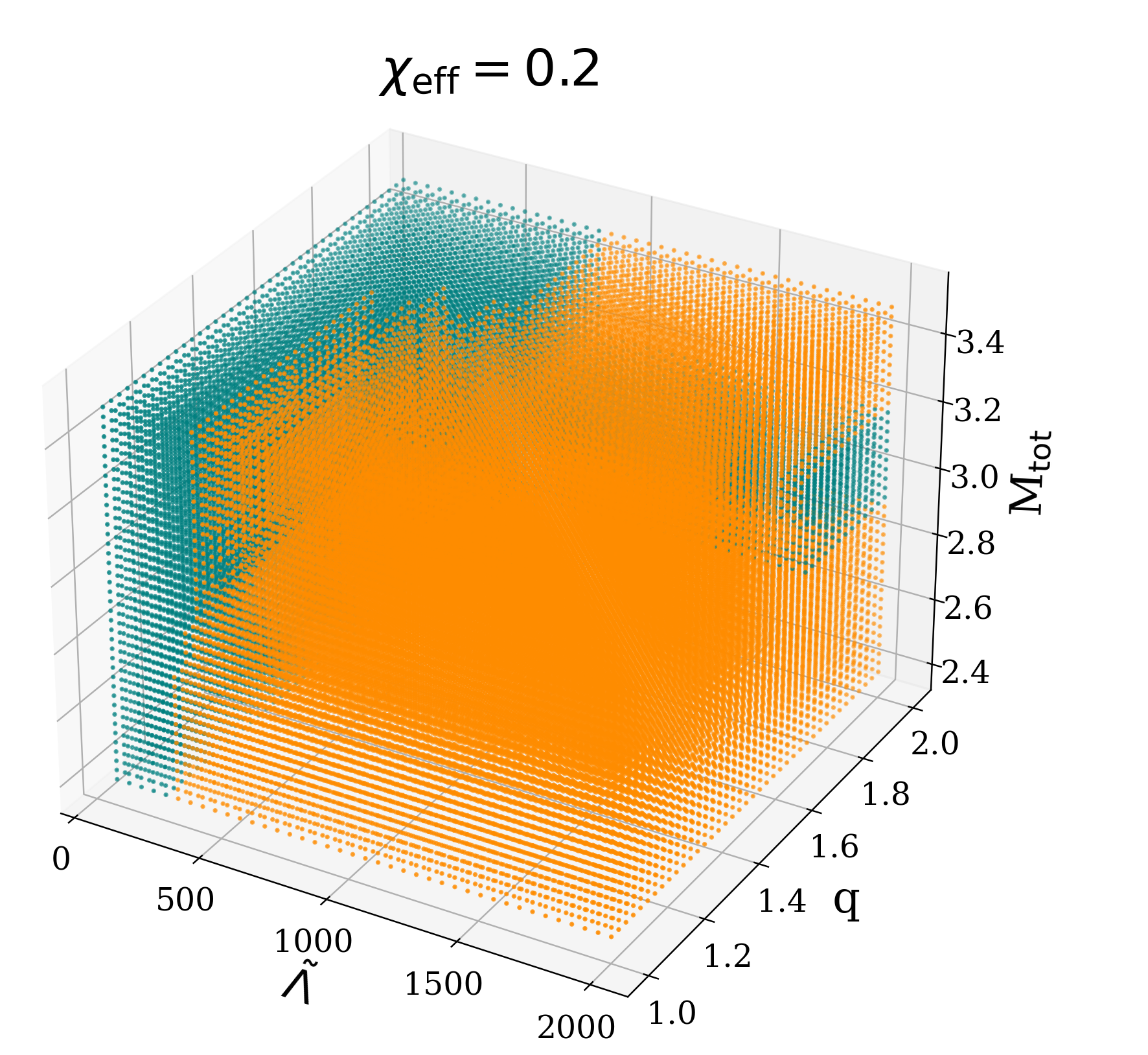}
        \includegraphics[width=0.32\textwidth]{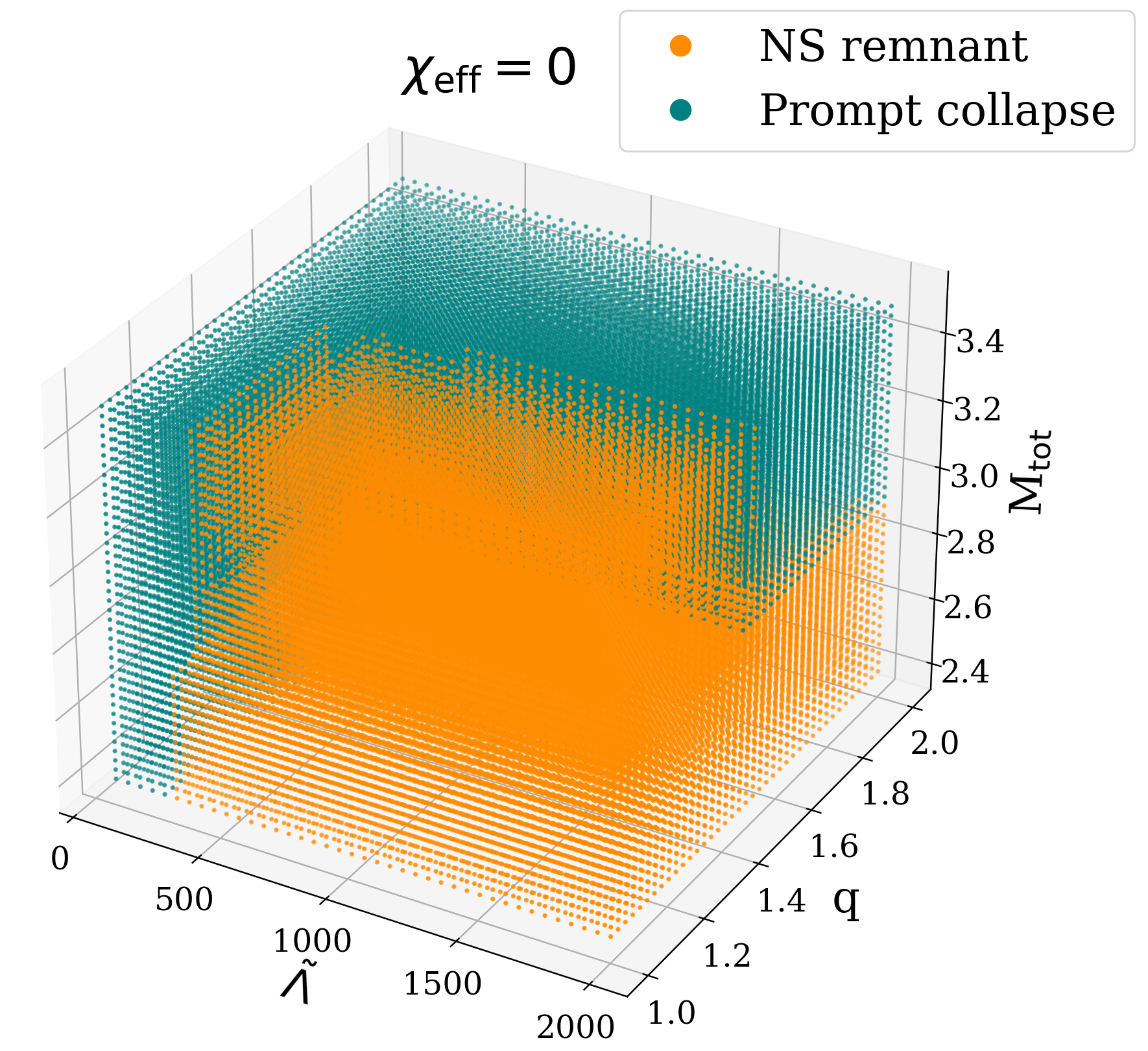}
        \includegraphics[width=0.32\textwidth]{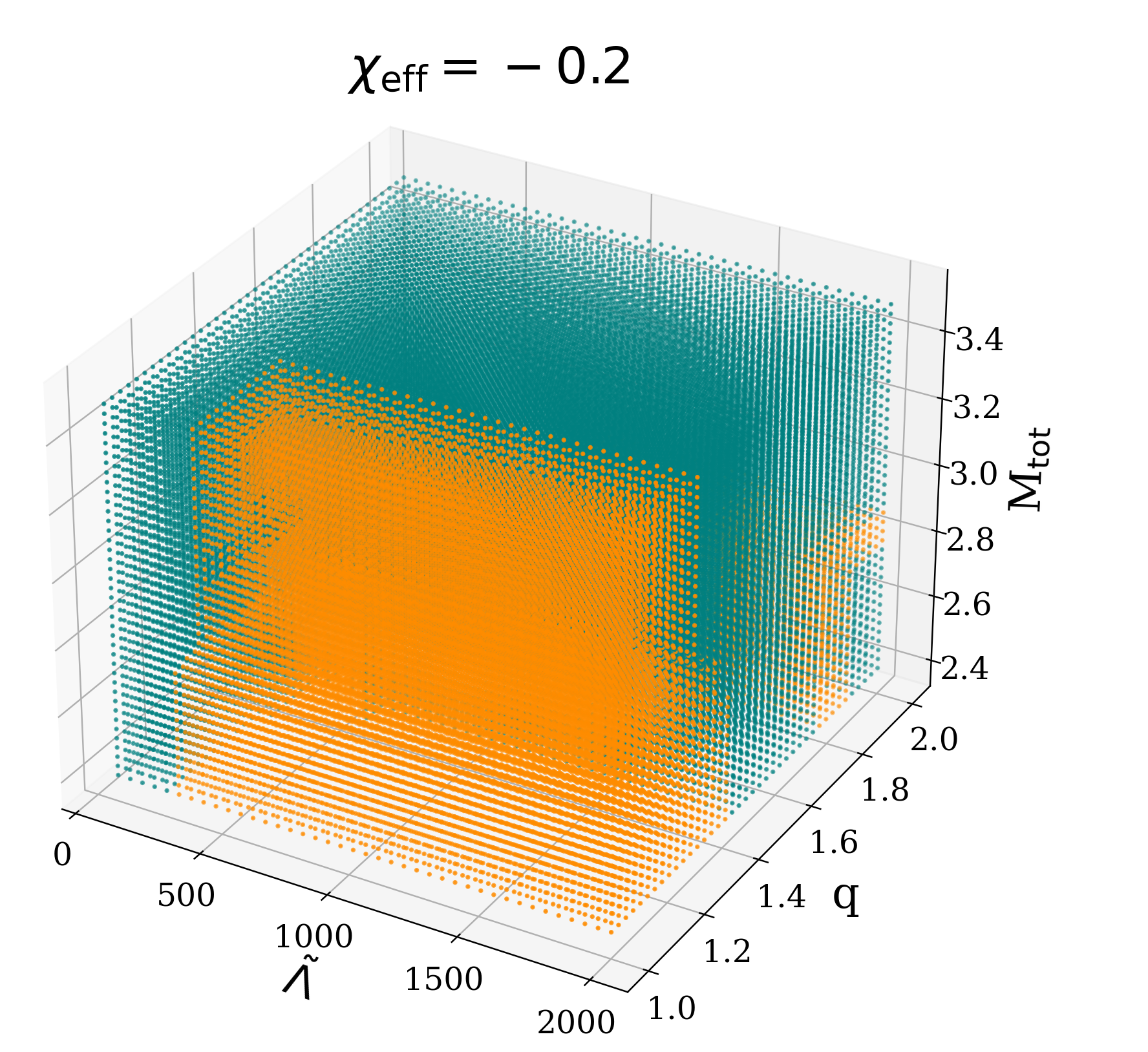}
    \caption{Outcomes predicted by Classifier A (orange for formation of a NS remnant, teal for prompt collapse) for a grid of points in the parameter space $M_{\rm tot}$, $q$, and $\tilde{\Lambda}$, for $\chi_{\rm eff} = 0.2$ (left panel), $\chi_{\rm eff} = 0.0$ (central panel), and $\chi_{\rm eff} = -0.2$ (right panel). }
    \label{fig:spin_distr}
\end{figure*}

\subsubsection{Prompt-collapse threshold and effect of parameters}

Because of the impact on possible electromagnetic counterparts, it is particularly important to predict whether a BNS system underwent prompt collapse or not. As mentioned in Sec.~\ref{sec:intro}, if the mass of the remnant exceeds the maximum mass supported with differential rotation, the system is bound to endure a prompt collapse. This mass threshold value $M_{\rm thr}$ is therefore considered the key parameter to determine whether the system forms a NS remnant or not. However, $M_{\rm thr}$ depends on the EoS, thus finding a connection between the remnant mass and the probability of prompt collapse is not straightforward. Refs.~\cite{Hotokezaka:2011dh,Bauswein:2013jpa} noticed that $M_{\rm thr}$ can be expressed as a function of $M_{\rm \textsc{tov}}$ as $M_{\rm thr} = \kappa \cdot M_{\rm \textsc{tov}}$, where $\kappa$ depends on the EoS. Subsequent studies improved this relation, taking into account also the effects of mass ratio and spins~\cite{Kolsch:2021lub,Kashyap:2021wzs,Agathos:2019sah,Bauswein:2020xlt,Perego:2021mkd,Tootle:2021umi,Koppel:2019pys}.

In our models, we do not include information about $M_{\rm \textsc{tov}}$ since it is not one of the parameters recovered from inspiral-GWs analysis. However, we look at the possible presence of threshold values for the prompt collapse in our features. Figure~\ref{fig:mass_lambda_gw170817} shows the distribution of $M_{\rm tot}$ and $\tilde{\Lambda}$ values for the posterior samples of GW170817+EoS, color-coded based on the different postmerger scenario predicted by Classifier A. A threshold in $\tilde{\Lambda}$ values clearly appears between samples that predict prompt collapse and the ones that predict the formation of a NS remnant. For simplicity, we do not show the dependence on $q$ or $\chi_{\rm eff}$, since $M_{\rm tot}$ and $\tilde{\Lambda}$ were found to be the dominant features in the classifier. If we fix $M_{\rm tot}$, $q$, and $\chi_{\rm eff}$ to the median values of the samples, $M_{\rm tot} = 2.735 \, M_\odot$, $q=1.353$, and $\chi_{\rm eff} =0.002$, we find the threshold value of $\tilde{\Lambda}$ to be
\begin{equation}
    \tilde{\Lambda}_{\rm th} \sim 310,
\end{equation}
slightly smaller than the value found in Ref.~\cite{Agathos:2019sah}. 

In order to understand the effect of mass ratio, we generate a grid of points with $M_{\rm tot} \in [2.4,3.5]\, M_\odot$, $q \in [1.,2.]$, and $\tilde{\Lambda} \in [85,2000]$. Although some of the points in this grid, for larger $M_{\rm tot}$ and lower $\tilde{\Lambda}$, are slightly outside the training-set range (see Tab.~\ref{tab:ranges}), here we intend to show a general qualitative behavior, and we do not expect the overall trends, i.e., larger values of $M_{\rm tot}$ and smaller values of $\tilde{\Lambda}$ leading more likely to a prompt collapse, to change over these slightly wider ranges. 

The central panel of Fig.~\ref{fig:spin_distr} shows the distribution of prompt collapse and NS remnant predictions, assuming $\chi_{\rm eff} = 0.0$. Besides the strong dependence on $\tilde{\Lambda}$, one can also distinguish the influence on $M_{\rm tot}$ and $q$, with higher values of both leading more likely to prompt collapse.
While it is expected that larger values of total mass lead to a prompt collapse, understanding the effects of mass ratio is not straightforward. However, different studies showed that mass asymmetry decreases $M_{\rm thr}$ for most EoSs\footnote{Ref.~\cite{Perego:2021mkd} pointed out that, for some specific EoSs, intermediate values of $q$ can lead to the opposite effect.} (see, e.g., Refs.~\cite{Kashyap:2021wzs,Tootle:2021umi,Kolsch:2021lub,Bauswein:2020xlt}), consistent with our model's behavior.

The left and right panels of Fig.~\ref{fig:spin_distr} show the distribution of remnant predictions for the same grid of parameters, but assuming $\chi_{\rm eff} = 0.2$ and $\chi_{\rm eff} = -0.2$, respectively. Although more NR data would be needed to find a precise relation, the case with positive $\chi_{\rm eff}$ clearly predicts less prompt collapses than the irrotational case, while negative spin results in more prompt collapse scenarios, in agreement with the findings in Ref.~\cite{Tootle:2021umi}.

\begin{figure*}[htb]
    \begin{minipage}[]{.48\textwidth}
        \centering
        \includegraphics[width=\textwidth]{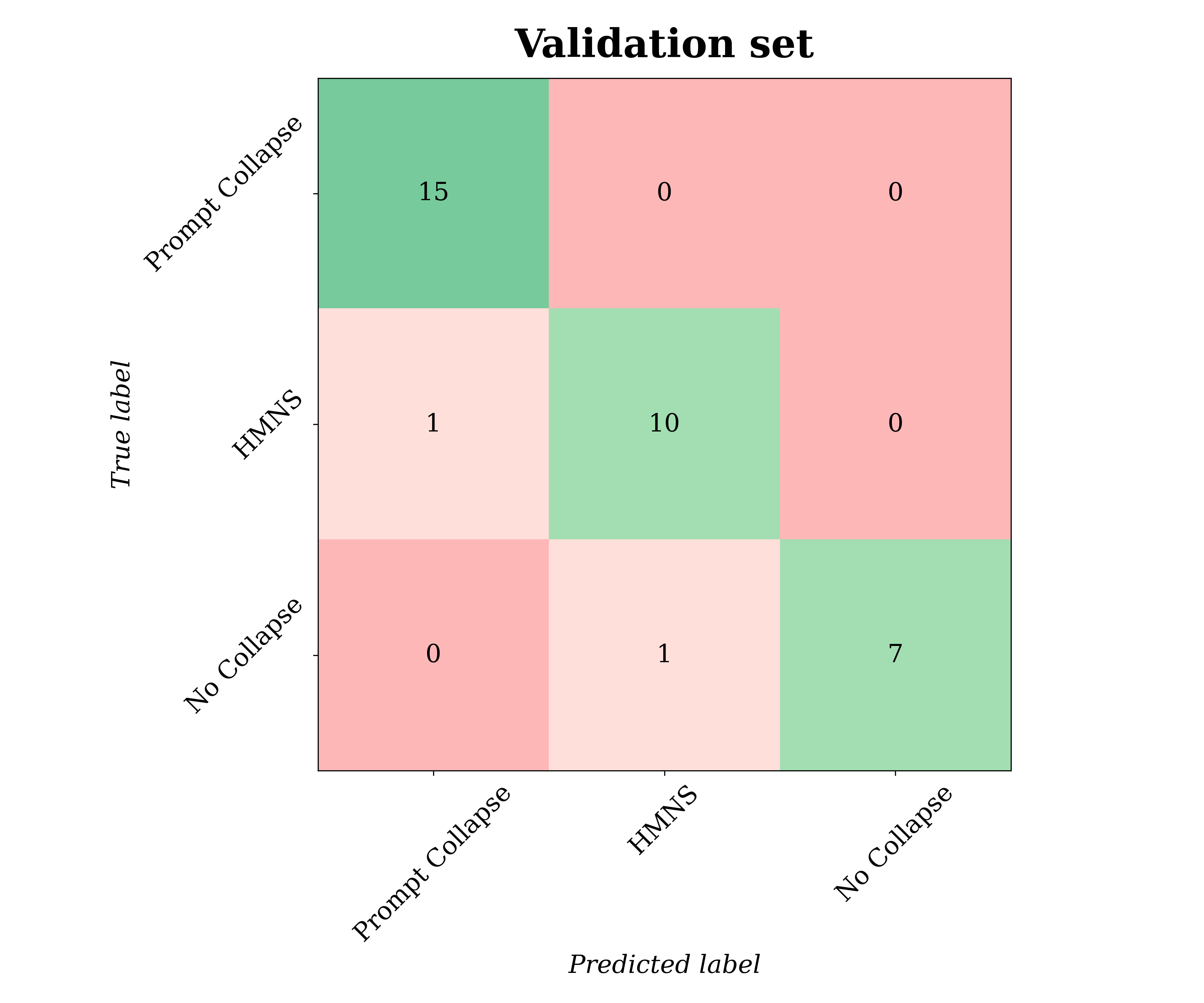}
    \end{minipage}
    \hfill
    \begin{minipage}[]{.48\textwidth}
        \centering
        \includegraphics[width=\textwidth]{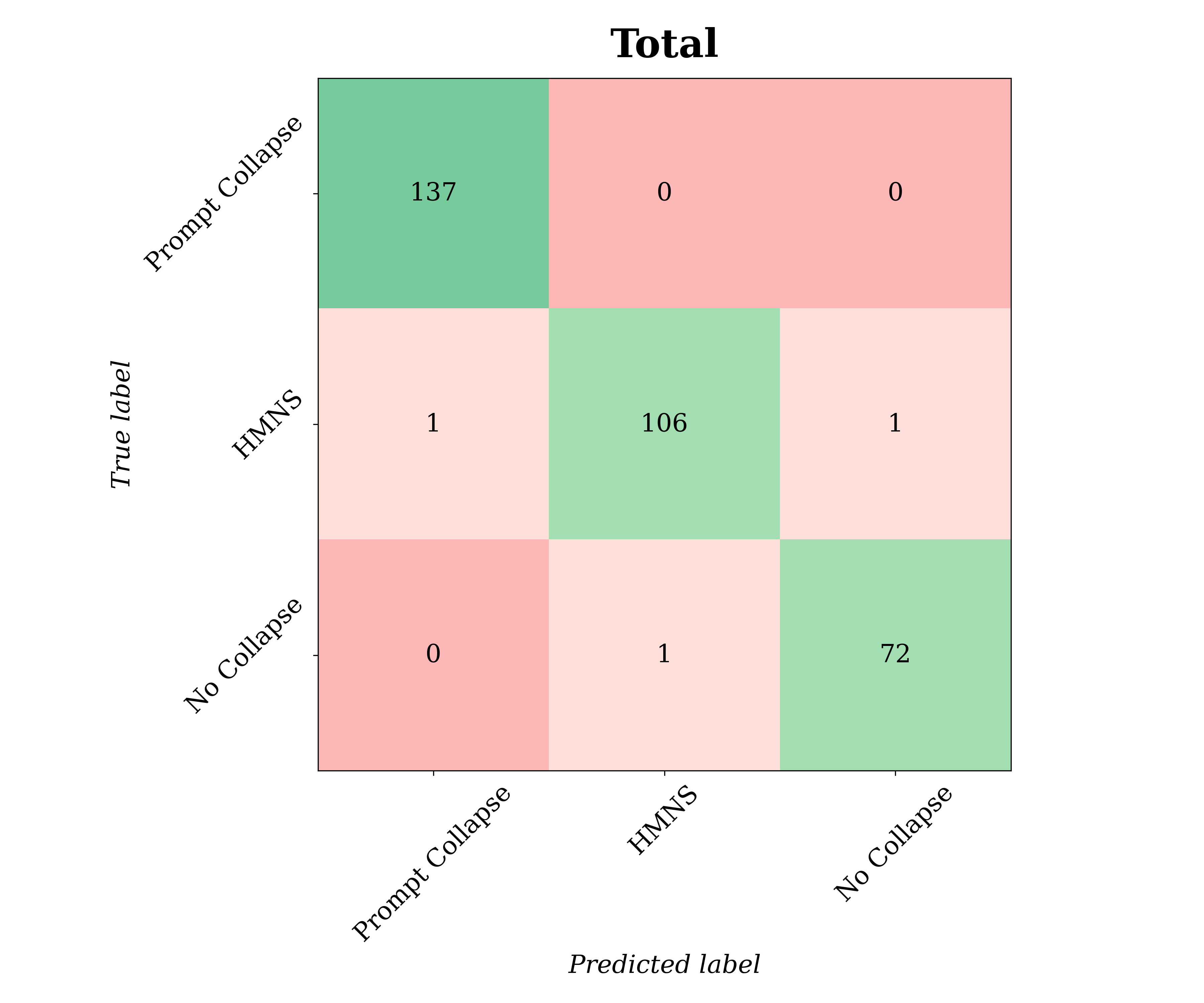}
    \end{minipage}  
    \caption{Confusion matrix for Classifier B, computed on the validation set (left panel) and on the complete dataset (right panel). The green color on the diagonal line highlights the entries that are correctly classified, while the red corresponds to the matrix elements with potential misclassifications.}
    \label{fig:confmatr_int}
\end{figure*}

\subsubsection{Application to low-latency data}
\label{sssec:lowlat}

As mentioned, one of the main applications of our classifier is to inform potential electromagnetic follow-up campaigns, especially with next-generation detectors, when we expect to observe a large number of BNS merger events. Currently, low-latency analyses provide point estimates of parameters obtained through template-based, matched-filtering searches. These estimates include the values of masses and spins, but not tidal deformabilities\footnote{At present, the templates employed in BNS searches do not incorporate tidal deformability because it would imply a large additional computational cost, while, given its relatively small effect on the overall phase, it is not expected to significantly improve the search results, and therefore binary black hole waveforms are deemed sufficient.}. Although this might change for future-generation detectors, since their increased sensitivity will require more and more accurate templates, for the moment we can marginalize over EoSs to estimate the value of $\tilde{\Lambda}$ from low-latency data (recently, EoS marginalization has been employed also for the prediction of kilonova curves in Ref.~\cite{Toivonen:2024ike}).
As an example, let us investigate what we would have obtained with the real events GW170817 and GW190425. For the point estimates of masses and spins released by the LVK public alerts (as reported in Table 1 in Ref.~\cite{Stachie:2021noh}), we consider the set of $10^5$ EoSs analyzed in Ref.~\cite{Koehn:2024set}, and for each one of them we compute the $\tilde{\Lambda}$ value corresponding to the low-latency mass estimates. 
At this point, we can either (i) take the $\tilde{\Lambda}$ estimate as their weighted mean, where the weights correspond to each EoS posterior probability as obtained through the analysis in Ref.~\cite{Koehn:2024set}\footnote{The posterior probabilities are provided by the web-interface in Ref.~\cite{eos_tool}. In particular, we chose a conservative estimate that included only information from the Black Widow J0952-0607, NICERJ0030+0451, NICERJ0740+6620, and GW170817 measurements; see Ref.~\cite{Koehn:2024set} for details.}; (ii) since the classifiers predict the probability for each class, and then choose as the output label the one with the highest probability, we can directly weigh the classes' probabilities. With both these methods, for GW170817 Classifier A predicts the formation of a NS remnant, while for GW190425 a prompt collapse to BH.

Both these approaches provide us with a prediction in a few seconds. Moreover, a lot of effort is ongoing to develop methods to perform fast parameter estimation analyses (see, e.g., Refs.~\cite{Dax:2021tsq, Dax:2024mcn, Wouters:2024oxj}), which are expected to deliver posterior samples within minutes, and hence, in the future, can be employed in the low-latency online analyses.

\subsection{Classifier B}
\label{ssec:cl_b}

Next, we build a classifier to distinguish between three different scenarios: prompt collapse to BH (with probability again dubbed $p_{\rm \textsc{pcbh}}$), formation of a HMNS that collapses within the simulation (with probability $p_{\rm \textsc{hmns}}$), or formation of a remnant that does not collapse within the simulation time ($p_{\rm \textsc{nc}}$). As explained in Sec.~\ref{ssec:dataset}, our dataset includes simulations whose remnant time, $\tau_{\rm BH}$, is registered as ``greater'' than the remaining simulation time after the merger, $t_{\rm sim}$. However, for some of them, this remaining time is very short, of the order of a few milliseconds. These cases are not informative, since a configuration for which we only know that the collapse time is larger than 5~ms, for example, could both be a HMNS or belong to class IV. Therefore, for Classifier B, as well as for Classifier C (see next Sec.~\ref{ssec:cl_c}), we remove from the dataset those points with $\tau_{\rm BH} > t_{\rm sim}$, if $t_{\rm sim} < 25$~ms. This new dataset comprises 318 configurations. 

As for Classifier A, the dataset is split into 90\% for training and 10\% for validation (see Sec.~\ref{ssec:data_split}), and we optimize the model's hyperparameters. The confusion matrices computed on the validation and complete set are shown in Fig.~\ref{fig:confmatr_int}, while the classifier's accuracy and $MCC$ are reported in Table~\ref{tab:accuracies}. Although both $\alpha$ and $MCC$ are slightly lower than the values found for Classifier A, as expected given the increased complexity of the task, they are still large, above $90\%$. 

We apply Classifier B to the real events GW170817 and GW190425, concluding that the most likely scenarios are that GW170817 produced a HMNS, while GW190425 resulted in a prompt collapse (see Table~\ref{tab:realev}). Also in this case, the more information is included when inferring the source parameters (see Ref.~\cite{Koehn:2024set} for details), the stronger the preference for one class. For GW190425, when using the posterior samples obtained from GW data only, the classifier predicts a non-negligible probability $p_{\rm \textsc{nc}}$ that the system outcome belongs to class IV, which is highly unlikely given the large total mass of the event. This is due to the fact that, in our dataset, most of the high-mass configurations have very stiff EoSs, such as MS1b or MS1~\cite{Read:2008iy}, with large $M_{\rm \textsc{tov}}$, and thus usually produce long-lived or stable remnants. Although these EoSs were likely ruled out by the GW170817 observation~\cite{LIGOScientific:2018hze}, we include these simulations in the dataset in order to not make any a-priori assumptions on the data or EoS. Nonetheless, when we consider the GW190425+EoS data, this trend disappears, and we obtain $p_{\rm \textsc{pcbh}} \sim 99\%$. 
For GW170817, when using the classifier on the samples obtained including information about both the EoS and the EM counterpart, the predicted probability for the formation of a HMNS reaches $p_{\rm \textsc{hmns}}=99.5\%$.

Also for Classifier B, we find that the feature with the strongest influence is $\tilde{\Lambda}$, followed by $M_{\rm tot}$. 

\begin{figure*}[htb]
    \begin{minipage}[]{.48\textwidth}
        \centering
        \includegraphics[width=\textwidth]{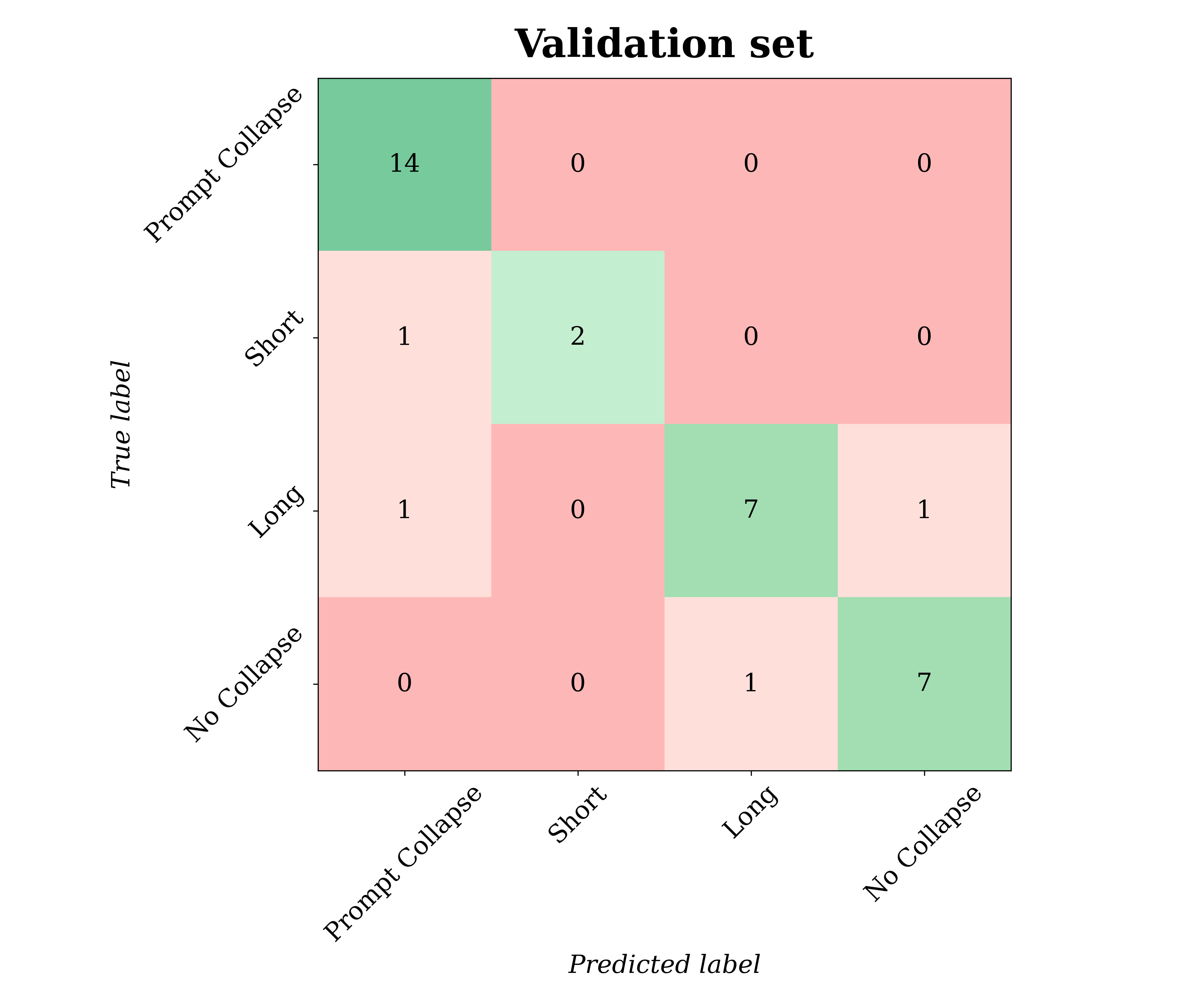}
    \end{minipage}
    \hfill
    \begin{minipage}[]{.48\textwidth}
        \centering
        \includegraphics[width=\textwidth]{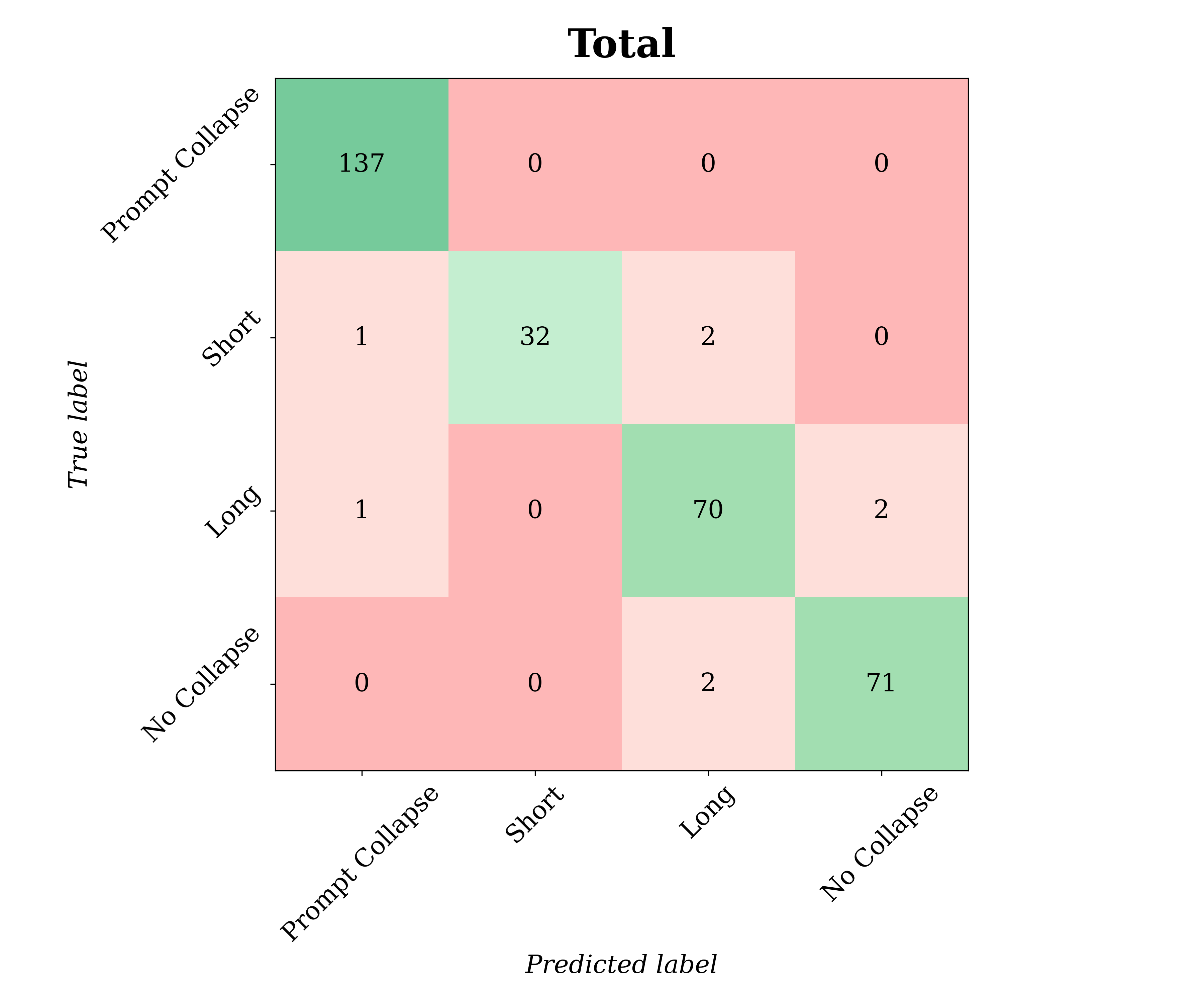}
    \end{minipage}  
    \caption{Confusion matrix for Classifier C, computed on the validation set (left panel) and on the complete dataset (right panel). The green color on the diagonal line highlights the entries that are correctly classified, while the red corresponds to the matrix elements with potential misclassifications.}
    \label{fig:confmatr_multi}
\end{figure*}

\subsection{Classifier C}
\label{ssec:cl_c}

Finally, we create a classifier to distinguish between all four classes, employing the modified dataset as for Classifier B. We follow the same procedure, first creating the training and validation set, as explained in Sec.~\ref{ssec:data_split}, and then optimizing the model hyperparameters. From Fig.~\ref{fig:confmatr_multi}, showing the confusion matrices obtained on the validation set and the complete dataset, and Tab.~\ref{tab:accuracies}, we see that the accuracy and $MCC$ are lower than the values found for the other two classifiers, as expected, but we still recover only a few misclassifications, with a validation accuracy $\sim 88\%$. In particular, we do not observe cases of ``extreme'' misclassification, i.e., in which a prompt collapse is classified as class IV or vice versa.

Studying the real events GW170817 and GW190425, we find that GW190425 resulted most likely in a prompt collapse to BH, while GW170817 produced a HMNS, with roughly the same probability of being a short-lived or long-lived HMNS (see Table~\ref{tab:realev}). For GW170817, analyzing the posteriors produced by GW data only, the classifier predicts $p_{\rm \textsc{pcbh}} = 41.7 \%$, $p_{\rm \textsc{long}} = 39.8 \%$, and  $p_{\rm \textsc{short}} = 15.6 \%$, with the latter two being the probability of a long-lived HMNS (case III) or short-lived HMNS (case II), respectively. Although the prompt collapse scenario yields the larger probability, one must take into account that the HMNS class in Classifier B is now split into $p_{\rm \textsc{short}}$ and $p_{\rm \textsc{long}}$, and their combined probability is larger than $p_{\rm \textsc{pcbh}}$. Again, if we apply the classifier to the samples obtained with additional information, the probability for the prompt collapse decreases. In particular, for GW170817+EoS+KN+GRB, Classifier C predicts $p_{\rm \textsc{pcbh}} =0.5\%$, $p_{\rm \textsc{short}} = 50.8\%$, $p_{\rm \textsc{long}}=48.6\%$, and $p_{\rm \textsc{nc}} = 0.1\%$.

Also in this case, we observe that the parameters with the larger impact in determining the fate of the system are $\tilde{\Lambda}$ ($\sim 0.50 $) and $M_{\rm tot}$ ($\sim 0.25$), with the latter one having more influence with respect to the case of Classifier A.


\section{Uncertainty in models' predictions}
\label{sec:uncertainty}

In this section, we discuss how to quantify the uncertainty in our classifiers' predictions. 

For each input point $\mathbf{x}$, the classifier computes the probabilities for $\mathbf{x}$ to belong to each class $\omega_k$, $P(\omega_k | \mathbf{x})$, and its final prediction is that $\mathbf{x}$ belongs to the class with the largest probability. The likelihood of the preferred class provides the confidence, and can be used as a measure of the uncertainty of the prediction~\cite{uncert_thesis}.  Figure~\ref{fig:pred_probs} shows the distribution of the prediction confidence for Classifier A over the GW170817 samples. Regarding as an overall uncertainty estimate the mean of this distribution, and considering, for example, the results of Sec.~\ref{ssec:cl_a}, we find that our model predicts GW170817 to have $60\%$ probability of formation of a NS remnant and $40\%$ prompt collapse with confidence $> 99\%$. For the GW170817 samples obtained with the additional EoS and EM information (GW170817, GW170817+EoS, GW170817+EoS+KN, and GW170817+EoS+KN+GRB), we obtain a prediction confidence $\sim 99\%$ for Classifier A, $\sim 94\%$ for Classifier B, and $83-90\%$ for Classifier C. This estimate of uncertainty basically corresponds to quantifying ``how sure'' the classifier is about its prediction; therefore if two or more classes have very similar probabilities the uncertainty will be larger. This explains why Classifier B and Classifier C yield larger prediction uncertainties, especially in the latter case, where the classifier predicts basically the same probability for case II and case III for GW170817+EoS+KN+GRB (see Table~\ref{tab:realev}).

For GW190425, the prediction confidence is larger than $90\%$ for all the classifiers, and it improves up to $\sim 99\%$ for predictions on the GW190425+EoS data.

It is important to highlight that the largest source of uncertainty comes from the NR simulations employed to train the models. Different physics (e.g., neutrino treatments), settings, and resolutions can produce different $\tau_{\rm BH}$ even if the initial system configuration is the same.\footnote{Examples of the effects of NR uncertainty can be found in Ref.~\cite{more_nr}, where the tables in Appendix C include the remnant classification obtained for different resolutions of the simulations. Additionally, Ref.~\cite{Nedora:2020hxc}, whose simulations are part of the CoRe database, shows, in Table II, the different collapse times obtained with different resolutions, and with or without simulating the effects of large magnetic fields with a subgrid model.}

In the event of a detection, we need to know how much we can trust the prediction of these classifiers, especially if it informs EM follow-up. 
In this case, the input $\mathbf{x}$ consists of only one single point with the estimated parameter values provided by low-latency analyses, and not the full posterior obtained through the subsequent parameter estimation process. 
Therefore, apart from estimating the confidence of the prediction, as explained above, we can compute the probability of that point being misclassified to further decide whether to trust the prediction or not. 
We can calculate the uncertainty of the prediction for a point $\mathbf{x}$ based on the prediction entropy, and compare this value with a threshold found for the specific classifier employed: if the prediction uncertainty for $\mathbf{x}$ is smaller than the threshold, we trust the classification; if, on the other hand, the uncertainty is greater than the threshold, the probability of the prediction being a misclassification is non-negligible, and, therefore, more investigations about the event are necessary. In particular, for Classifiers A, B, and C, we find threshold values $T_{\rm \textsc{a}} \sim 0.3$, $T_{\rm \textsc{b}} \sim 0.06$, and $T_{\rm \textsc{c}} \sim 0.09$, respectively. The details about the uncertainty and the threshold calculations are explained in Appendix~\ref{sec:appendix_unc}.

\begin{figure}[htb]
        \centering
        \includegraphics[width=\linewidth]{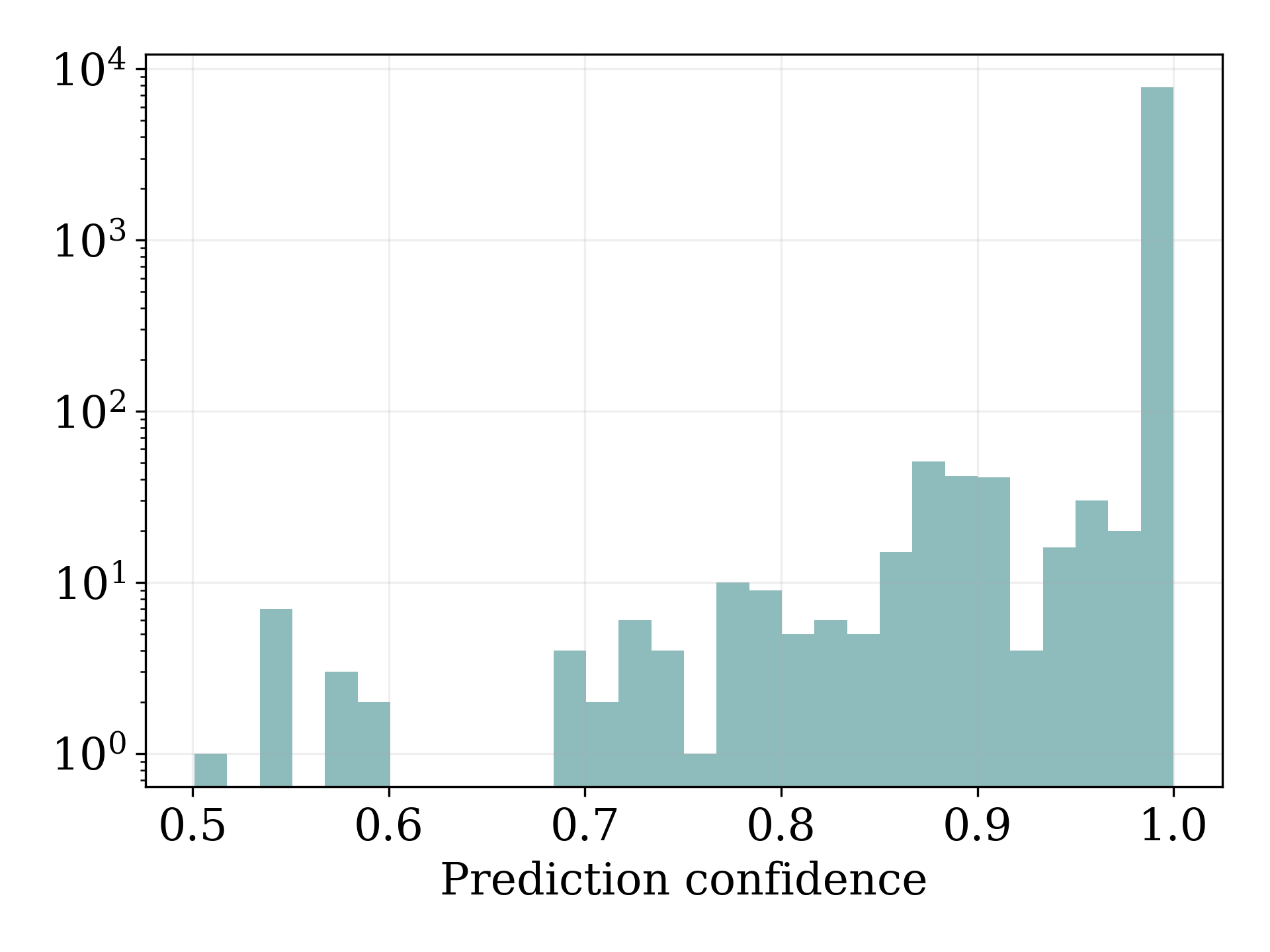}
    \caption{Distribution of confidence for the predictions of Classifier A on the GW170817 samples. Note that the y-axis is shown with a logarithmic scale.}
    \label{fig:pred_probs}
\end{figure}

\section{Conclusions}
\label{sec:conclusions}

We built a classifier to predict the outcome of a BNS merger based on parameters that can be measured from the inspiral GW signal, $M_{\rm tot}$, $q$, $\tilde{\Lambda}$, and $\chi_{\rm eff}$. We employed the Gradient Boosted Decision Tree algorithm in \texttt{sklearn} and a dataset comprising publicly available data from NR simulations.
We constructed and trained three different models: Classifier A, to distinguish between prompt collapse to BH and formation of a NS remnant; Classifier B, to discriminate between prompt collapse, formation of a HMNS that collapses during the simulation, and formation of a NS remnant that does not collapse within the simulation time; Classifier C, which further discerns between short-lived ($2 \, {\rm ms} < \tau_{\rm BH} < 5 \, {\rm ms}$) and long-lived ($\tau_{\rm BH} > 5 \, {\rm ms}$) HMNSs.
All of them yielded very high values for both accuracy and the Matthews correlation coefficient (see Table~\ref{tab:accuracies}), where the latter takes into account the distribution of correct classifications among the different classes. The lowest values obtained were $\alpha = 88.2\%$ and $MCC = 0.831$ for Classifier C, as expected given its more complex task.

We investigated the outcome of GW170817 and GW190425 with the three classifiers, supplying as input the samples for $M_{\rm tot}$, $q$, $\tilde{\Lambda}$, and $\chi_{\rm eff}$ produced by parameter estimation analysis on the GW data. We concluded that GW170817 had a $\sim 60\%$ probability of forming a NS remnant, most likely a HMNS that collapsed in a few ms, while GW190425 collapsed promptly to a BH, with a $60-70 \%$ probability. Moreover, we also applied our classifiers to the posterior samples obtained in Ref.~\cite{Koehn:2024set}, including in the inference information about the EoS and, in the case of GW170817, the EM counterpart. In this case, the probability for GW170817 to have formed a HMNS  becomes larger than $99\%$, with roughly the same probability to be short- and long-lived, while GW190425 has more than $98\%$ probability to have undergone prompt collapse (see Table~\ref{tab:realev} for details). 

For all three classifiers, the most significant feature in determining the prediction is $\tilde{\Lambda}$, as expected since it is the parameter encoding the information about the EoS. Although the importance of $\chi_{\rm eff}$ is overall quite low, for Classifier A we find that, in general, binaries with $\chi_{\rm eff} < 0$ ($\chi_{\rm eff}>0$) are more likely to produce a prompt collapse to BH (NS remnant) with respect to the irrotational case (see Fig.~\ref{fig:spin_distr}). However, only a few of the simulations included in our dataset corresponded to spinning configurations, and we did not consider misaligned spins, thus more data are needed to systematically investigate the spin influence.

For Classifier A, which discriminates between prompt collapse and formation of a NS remnant, in the case of the GW170817 event we observed a clear threshold value for $\tilde{\Lambda}$ between the two scenarios (see Fig.~\ref{fig:mass_lambda_gw170817}). Specifically, for the samples median values of $M_{\rm tot}$, $q$, and $\chi_{\rm eff}$, we find that prompt collapse occurs for  $\tilde{\Lambda} < \tilde{\Lambda}_{\rm th} \sim 310$.

Finally, we discussed the uncertainty in the predictions of our models based on the probability assigned to each class, showing that our results on the real events data yielded a prediction confidence of roughly $90\%$ or higher.

Knowing the remnant produced by a BNS merger could provide us with information about the EoS of NSs' supranuclear-dense matter, the physical processes involved in the postmerger phase, and possible EM counterparts. Moreover, predicting whether a BNS system undergoes a prompt collapse to BH or produces a NS remnant is crucial to build postmerger models and to inform EM follow-up campaigns, especially for 3G detectors, for which a large number of BNS detections are expected. The classifiers developed in this work 
can be employed with any BNS detection, without the need for observing a postmerger signal. In fact, due to the high frequencies involved, the postmerger part of the GW signal is difficult to detect, and we expect to start observing it only with 3G detectors, but also in that case it could be weak~\cite{Puecher:2022oiz,Branchesi:2023mws}. On the other hand, our models enable us to predict the merger outcome also for potential detections with Advanced Virgo and Advanced LIGO; moreover, with future-generation detectors, the more precise measurements of the binary parameters from the inspiral signal will allow us to get even more precise predictions, even without including additional information.

\acknowledgements
We thank Chris Van Den Broeck, Brian Healy, Thibeau Wouters, and Vsevolod Nedora for the helpful discussion. The authors would like to thank Sebastian Khan for suggesting the use of interpretable machine learning techniques, such as SHAP, to enhance the understanding of our model and analysis.
The authors acknowledge funding from the Daimler and Benz Foundation for the project “NUMANJI” and from the European Union (ERC, SMArt, 101076369). Views and opinions expressed are those of the authors only and do not necessarily reflect those of the European Union or the European Research Council. Neither the European Union nor the granting authority can be held responsible for them. 
This material is based upon work supported by NSF’s LIGO Laboratory which is a major facility fully funded by the National Science Foundation, as well as the Science and Technology Facilities Council (STFC) of the United Kingdom, the Max-Planck-Society (MPS), and the State of Niedersachsen/Germany for support of the construction of Advanced LIGO and construction and operation of the GEO600 detector. Additional support for Advanced LIGO was provided by the Australian Research Council. Virgo is funded, through the European Gravitational Observatory (EGO), by the French Centre National de Recherche Scientifique (CNRS), the Italian Istituto Nazionale di Fisica Nucleare (INFN) and the Dutch Nikhef, with contributions by institutions from Belgium, Germany, Greece, Hungary, Ireland, Japan, Monaco, Poland, Portugal, Spain. KAGRA is supported by Ministry of Education, Culture, Sports, Science and Technology (MEXT), Japan Society for the Promotion of Science (JSPS) in Japan; National Research Foundation (NRF) and Ministry of Science and ICT (MSIT) in Korea; Academia Sinica (AS) and National Science and Technology Council (NSTC) in Taiwan.

\appendix

\section{Assessing the probability of misclassification}
\label{sec:appendix_unc}

In general, the uncertainty of predictions obtained through machine learning algorithms are divided into two classes~\cite{uncert_thesis}:
\begin{itemize}
    \item \emph{Irreducible} or \emph{data uncertainty}, caused by the complexity of the data, possible multimodal features or noise. In classification problems, it is defined as the entropy of the conditional probability of a given input $\mathbf{x}$ belonging to a class $k$
    \begin{equation}
        \mathcal{H}[p(y|\mathbf{x})] = -\sum_{k=1}^K p(y=\omega_k | \mathbf{x}) \ln p(y=\omega_k | \mathbf{x}),
        \label{eq:data_unc}
    \end{equation}
    where $\left\{\mathbf{x}\right\}$ is the set of inputs, $K$ is the total number of classes, and $y=\left\{\omega_1, ..., \omega_K \right\}$ are the labels. 
    \item \emph{Knowledge} or \emph{epistemic uncertainty}, which derives from the model lacking the capability to describe (some) data (this happens, for example, when the input for which we want the model's prediction comes from a different distribution than the one over which the model was trained).
\end{itemize}

The \emph{knowledge uncertainty} quantifies potential shortcomings of the model, but it cannot be directly computed from the model's predictions. To assess such model uncertainty, we follow the \emph{Bayesian ensemble-based framework} described in Refs.~\cite{2020arXiv200610562M,uncert_thesis}. Basically, we consider an ensemble of models and their different predictions on a given dataset. In this context, we can estimate the \emph{knowledge uncertainty} as the spread of predictions across the models ensemble. More specifically, we can compute it as the difference between the total uncertainty and the expected data uncertainty, considering the mutual information between the models parameters $\mathbf{\theta}$ and the predictions $y$~\cite{2020arXiv200610562M,2017arXiv171007283D}
\small
\begin{equation}
\mathcal{I}[y,\mathbf{\theta}] \sim \mathcal{H} \left[ \frac{1}{M} \sum_{m=1}^M p(y|\mathbf{x};\mathbf{\theta}^{\left(m\right)})\right] - \frac{1}{M} \sum_{m=1}^M \mathcal{H} \left[  p(y|\mathbf{x};\mathbf{\theta}^{\left(m\right)})\right],
\label{eq:know_unc}
\end{equation}
\normalsize
where each model $m$ has parameters $\mathbf{\theta}^{\left( m \right)}$, and $M$ is the total number of models.

To generate the models ensemble, we employ the \textit{stochastic gradient boosting} method~\cite{FRIEDMAN2002367} in \texttt{sklearn} with different random states. The method illustrated in Ref.~\cite{2020arXiv200610562M} proposes more sophisticated ways to generate the ensemble, which, however, imply a higher computational cost; given that, in any case, our uncertainties are dominated by the NR errors, we deem stochastic gradient boosting sufficient. 

In the following, we show how we can derive the \emph{knowledge uncertainty} with the \emph{Bayesian ensemble} method, considering as an example Classifier A. We generate an ensemble of 100 models and compute the knowledge uncertainty $\mathcal{U}(\mathbf{x})$ on each point $\mathbf{x}$ in the validation set as described in Eq.~\ref{eq:know_unc}. The underlying idea is that larger (smaller) values of $\mathcal{U}(\mathbf{x})$ indicate that the prediction for the point $\mathbf{x}$ is more (less) likely to be a misclassification. Therefore, following Ref.~\cite{uncert_thesis}, we can use the uncertainty $\mathcal{U}(\mathbf{x})$ to consider a point $\mathbf{x}$ correctly classified (label 0) or, instead, misclassified (label 1), based on a threshold $T$:
\begin{equation}
    \mathcal{I}_T(\mathbf{x}) = \begin{cases}
                                1, & \mathcal{U}(\mathbf{x}) > T, \\
                                0, & \mathcal{U}(\mathbf{x}) \le T.
                                \end{cases}
\label{eq:thresh_unc}
\end{equation}

Looking at our validation set, for which we know if the points are correctly classified or not, we check if, for different thresholds $T$, the criterion in Eq.~\ref{eq:thresh_unc} would have correctly identified them as misclassified or not, and we compute the area under the Receiver-Operating-Characteristic curve (AUROC)~\cite{FAWCETT2006861}. Since ROC curves include only information about true and false positive rates, the AUROC values can be biased toward 1.0 if the dataset includes significantly more negatives than positives. Therefore, we also calculate the value for the area under the precision-recall curve (AUPR), where the recall is the true positive rate and the precision is the fraction of true positives compared to the total number of points labeled as positives~\cite{manning2008introduction,Flach2015PrecisionRecallGainCP}. For both AUROC and AUPR, the closer their value is to 1, the more robust the classifier is. For the example of Classifier A with an ensemble of 100 models, we find AUROC = 0.9875 and AUPR = 0.79.

In the event of a detection, therefore, we can evaluate whether the related prediction is likely a misclassification, and therefore more investigations are needed, by looking at its uncertainty $\mathcal{U}(x)$ in relation to an optimal threshold value. However, considering an ensemble of models every time is computationally expensive and unnecessarily elaborate given that, as already mentioned, NR errors dominate our uncertainties anyway. Therefore, we perform a similar check, but considering only one model, i.e., Classifier A, not an ensemble. We compute the data uncertainty on this single model as in Eq.~\ref{eq:data_unc}, and we build an ROC curve following the same procedure described above.
We look for the optimal threshold value $T_{\rm \textsc{a}}$, which keeps the false positive rate below 0.1 and the true positive rate close to 1, and we find $T_{\rm \textsc{a}} \sim 0.3$. 
With the same method, we look for the ideal threshold for Classifier B, $T_{\rm \textsc{b}}$. In this case, if we want to keep the false positive rate small ($\sim 0.1$), we need an uncertainty threshold 0.67, which corresponds to just above a $50\%$ true positive rate; if, instead, we want a true positive rate close to $100\%$, we need to set the uncertainty threshold to 0.06, which, however, implies a quite high false positive rate (however, such a conservative choice is acceptable since the goal of this uncertainty threshold is to identify inputs that would require more study). 
Finally, repeating the procedure for Classifier C, we conclude that setting a threshold $T_{\rm \textsc{c}}=0.09$ allows us to have the true positive rate basically at 1, while keeping the false positive rate $\sim 0.2$.

\bibliography{refs}

\end{document}